\documentclass[prb,
floatfix,
showpacs,
twocolumn,
floats,
10pt,
aps,
citeautoscript,
longbibliography,
superscriptaddress]{revtex4-2}

\usepackage{babel}
\usepackage{lmodern}
\usepackage{graphicx}
\usepackage{dcolumn}
\newcolumntype{d}[1]{D{.}{.}{#1}}
\usepackage{booktabs}
\usepackage{amsmath}
\usepackage{amssymb}
\usepackage{physics}
\usepackage{bbm}
\usepackage{accents}
\usepackage{scalefnt}

\newcommand{\Eq}[1]{Eq.~\eqref{#1}}
\newcommand{\Eqs}[1]{Eqs.~\eqref{#1}}

\newcommand{\Fig}[1]{Fig.~\ref{#1}}
\newcommand{\Figs}[1]{Figs.~\ref{#1}}
\newcommand{\Sec}[1]{Sec.~\ref{#1}}
\newcommand{\App}[1]{App.~\ref{#1}}
\newcommand{\Appendix}[1]{Appendix~\ref{#1}}
\newcommand{\Appendices}[1]{Appendices~\ref{#1}}

\newcommand{\I}{\mathrm{i}}
\DeclareMathOperator{\diag}{diag}

\newcommand{\kB}{k_\mathrm{B}}
\newcommand{\D}{\mathrm{d}}
\newcommand{\mH}{\mathrm{HF}}
\newcommand{\CR}[1]{a^{\dag}_{#1}}
\newcommand{\AN}[1]{a^{\phantom{\dag}}_{#1}}

\newcommand{\brakettt}[3]{\left\langle#1\middle|#2\middle|#3\right\rangle}

\newcommand{\TRI}[1]{\mathcal{#1}}
\newcommand{\AND}[1]{\mathtt{#1}}

\newcommand{\mr}[1]{\mathrm{#1}}
\newcommand{\mc}[1]{\mathcal{#1}}

\def\doubleC{{\mathbbm{C}}}
\def\doubleR{{\mathbbm{R}}}

\def\pdag{{\phantom{\dagger}}}

\def\HA{H_\mathrm{A}}
\def\Ht{H_\mathrm{t}}

\newcommand{\peakbar}[1]{
  \mathchoice
    {\accentset{\rule{0.7em}{0.4pt}}{#1}}
    {\accentset{\rule{0.7em}{0.4pt}}{#1}}
    {\accentset{\rule{0.65em}{0.3pt}}{#1}}
    {\accentset{\rule{0.45em}{0.25pt}}{#1}}
}

\def\bM{{\peakbar{M}}}
\def\bN{{\peakbar{N}}}
\def\bU{\peakbar{U}}
\def\bOmega{{\peakbar{\Omega}}}
\def\bara{{\bar{a}}}

\def\ta{\widetilde{a}}
\def\tA{\widetilde{A}}
\def\tB{\widetilde{B}}
\def\tG{\widetilde{G}}
\def\tH{\widetilde{H}}
\def\tN{\widetilde{N}}
\def\tOmega{\widetilde{\Omega}}
\def\tR{\widetilde{R}}
\def\tS{\widetilde{S}}
\def\tU{\widetilde{U}}

\newcommand{\SigmaPole}[1]{\alpha^\Sigma_{#1}}
\newcommand{\SigmaWeight}[1]{W^\Sigma_{#1}}
\newcommand{\SigmaHyb}[1]{B^\Sigma_{#1}}

\newcommand{\GPole}[1]{{\alpha_{#1}}}
\newcommand{\GWeight}[1]{{W_{#1}}}
\newcommand{\GHyb}[1]{{B_{#1}}}

\newcommand{\AugGPole}[1]{\widetilde{\alpha}_{#1}}
\newcommand{\AugGWeight}[1]{{\widetilde{W}_{#1}}}
\newcommand{\AugGHyb}[1]{\widetilde{B}_{#1}}

\newcommand{\HeffPole}[1]{\bar{\alpha}_{#1}}
\newcommand{\HeffWeight}[1]{{\peakbar{W}_{#1}}}
\newcommand{\HeffHyb}[1]{\peakbar{B}_{#1}}

\def\BlockVector{B}
\def\AugBlockVector{\widetilde{B}}
\def\OmegaSigma{\Omega^\Sigma}
\def\BlockVectorSigma{B^\Sigma}
\def\BlockVectorHeff{\peakbar{B}}
\def\KrylovState{K}

\def\MatElOne{\mu}
\def\MatElTwo{\nu}

\newenvironment{medpmatrix}{\scalefont{0.8}\begin{pmatrix}}{\end{pmatrix}}
\newenvironment{smallpmatrix}{\left(\begin{smallmatrix}}{\end{smallmatrix}\right)}

\newcommand{\mi}{\mathfrak{m}}

\newcommand{\LMUMunich}{Arnold Sommerfeld Center for Theoretical Physics, Center for NanoScience, and Munich Center for Quantum Science and Technology, Ludwig-Maximilians-Universit\"at M\"unchen, 80333 Munich, Germany}

\newcommand{\Heidelberg}{Institute for Theoretical Physics, Heidelberg University, Philosophenweg 19, 69120 Heidelberg, Germany}

\newcommand{\Vienna}{Institute of Solid State Physics, TU Wien, 1040 Vienna, Austria}

\newcommand{\Cologne}{Institute for Theoretical Physics, University of Cologne, 50937 Cologne, Germany}

\newcommand{\Rutgers}{Department of Physics and Astronomy and Center for Condensed Matter Theory, Rutgers University, Piscataway, New Jersey 08854-8019, USA}

\def\Quanty{\textsc{Quanty}}
\def\RAS{\texttt{RAS\_DMFT.jl}}

\RequirePackage[
  hyperindex,colorlinks,bookmarksnumbered,
  plainpages=true,pdfstartview=FitH]{hyperref}
\hypersetup{linkcolor=blue,urlcolor=blue,citecolor=blue}
\usepackage[all]{hypcap}
\usepackage{orcidlink}

\begin{document}

\preprint{}

\title{Symmetric estimator for discrete self-energy of discrete many-body systems}

\author{Aleksandrs Zacinskis\orcidlink{0000-0002-1780-3497}}
\affiliation{\Heidelberg}

\author{Frank T. Ebel\orcidlink{0009-0005-0479-0243}}
\affiliation{\Vienna}

\author{Mathias Pelz\orcidlink{0009-0001-3282-6742}}
\affiliation{\LMUMunich}

\author{Fabian B. Kugler\orcidlink{0000-0002-3108-6607}}
\affiliation{\Cologne}

\author{Karsten Held\orcidlink{0000-0001-5984-8549}}
\affiliation{\Vienna}

\author{Jan von Delft\orcidlink{0000-0002-8655-0999}}
\affiliation{\LMUMunich}

\author{Maurits W. Haverkort\orcidlink{0000-0002-7216-3146}}
\affiliation{\Heidelberg}

\author{Andreas Gleis\orcidlink{0000-0001-6260-5281}}
\affiliation{\Rutgers}

\date{\today}

\begin{abstract}

    We derive a discrete spectral representation of the single-particle self-energy using a discrete evaluation of Kugler's symmetric improved estimator. Our construction can be used on both the real and the complex (Matsubara) frequency axis. It is guaranteed to remain causal at the numerical level, in contrast to standard approaches that may generate unphysical negative spectral weight or require additional broadening.
    Our representation can be used for any Hamiltonian; here we apply it to
    quantum impurity models and in dynamical mean-field theory. The latter is formulated with a discrete hybridization function throughout its self-consistency loop. In both cases and across various numerical methods, we obtain significantly improved accuracy for a range of impurity properties.

\end{abstract}

\maketitle

\section{Introduction}

Quantum impurity models, introduced by Anderson in 1961~\cite{anderson_localized_1961}, feature a small number of interacting spin-orbitals that hybridize with a large set of non-interacting bath modes. Originally devised to explain the Kondo effect~\cite{kondo_resistance_1964},
they also --among others-- model  quantum dots~\cite{Kouwenhoven2001} and
are employed in dynamical mean-field theory (DMFT)~\cite{georges_dynamical_1996}
which maps an interacting lattice problem onto a self-consistently determined impurity model. Closely related, in quantum chemistry,
impurity models emerge in  quantum embedding approaches~\cite{potthoff_self-energy-functional_2003,huang_potential-functional_2011,wesolowski_frozen-density_2015, knizia_density_2012, lan_generalized_2017}.

To calculate impurity observables and spectral functions, different impurity solvers have been devised, ranging from perturbative and semi-analytical approaches~\cite{keiter_perturbation_1970, bickers_conserving_1989, pruschke_anderson_1989, hewson_kondo_1993, logan_local_1998} to quantum Monte Carlo (QMC) algorithms~\cite{hirsch_monte_1986, rubtsov_continuous-time_2005, werner_continuous-time_2006, gull_continuous-time_2011} to Hamiltonian-based methods with a discretized bath.
In the latter class, one approximates the bath spectral function by a finite set of poles and solves the resulting finite Anderson impurity Hamiltonian.
This is done, e.g., in exact diagonalization and Lanczos/Krylov approaches~\cite{caffarel_exact_1994, capone_solving_2007, weber_augmented_2012}, restricted active space solvers~\cite{gunnarsson_electron_1983, zgid_dynamical_2011, zgid_truncated_2012, haverkort_multiplet_2012,lu_efficient_2014, lu_exact_2017, lu_natural-orbital_2019}, numerical renormalization group (NRG)~\cite{wilson_renormalization_1975, krishna-murthy_renormalization-group_1980, bulla_numerical_2008, Weichselbaum2012b_QSpace}, and tensor-network approaches~\cite{nishimoto_dynamical_2006, weichselbaum_variational_2009, ganahl_efficient_2015, wolf_chebyshev_2014, wolf_imaginary-time_2015, bauernfeind_fork_2017, cao_tree_2021}.

In many cases, most notably DMFT, it is further important to reliably extract the single-particle  self-energy. A direct evaluation via the Dyson equation
\begin{align}
    \label{eq:SigmaG0G}
    \Sigma (z) =G_0^{-1}(z) - G^{-1}(z) \, ,
\end{align}
where
$G_0(z)$ and $G(z)$ are the Green functions of two different Hamiltonians, the non-interacting and the interacting one,
is ill-conditioned.
For a discretized bath, this is exacerbated as
the poles of $G_0(z)$ and $G(z)$
will be at different energies $z$. Even with a tiny numerical offset in the poles of $G(z)$ or numerical limitations in the number of poles, delicate cancellations are not perfect, and the self-energy becomes non-causal.

This can be avoided by an explicit broadening of the Green’s functions
or spectra~\cite{raas_spectral_2005, peters_spectral_2011, lee_adaptive_2016}, which can be further stabilized by  imposing exact high-frequency moments or asymptotic limits~\cite{wang_high-frequency_2011, li_efficient_2012} or by schemes that explicitly enforce analyticity and causality~\cite{han_causal_2021, labollita_stabilizing_2025}.
Alternatively, one can work with a  discrete pole representation~\cite{lu_efficient_2014}
and merge unphysical
negative spectral weight of $\Sigma(z)$ with positive weight so that causality is eventually achieved.

The numerical instabilities inherent to the subtraction
$G_0^{-1}(z)\!-\!G^{-1}(z)$ can be fully avoided
by using the equations of motion for the Green's function. This substantially more stable extraction scheme was introduced by Bulla, Hewson, and \mbox{Pruschke}~\cite{bulla_numerical_1998} for NRG, and later transferred to QMC~\cite{hafermann_improved_2012} and improved to a symmetric variant by Werner and Millis~\cite{werner_hybridization_2006} and Kaufmann {\em et al.}~\cite{kaufmann_symmetric_2019}. This, in turn, inspired the largely improved symmetric self-energy estimator by Kugler~\cite{kugler_improved_2022}. For real-frequency calculations, however,
again an artificial broadening is applied.

As a consequence,
a discrete spectral representation (DSR), i.e., a pole expansion for the self-energy, is not yet available. Access to such a DSR is desirable when computing quantities that are sensitive to broadening artifacts; e.g., the asymptotic low-energy behavior of the self-energy or the DMFT self-consistency loop.

In this paper, we show how to compute the discrete self-energy of discrete many-body systems in a manner that avoids delicate cancellations and involves no broadening. The key idea is to use Kugler's symmetric improved estimator for self-energies~\cite{kugler_improved_2022}, recognize that it corresponds to the Schur complement of a sub-block of an augmented Green's function, and compute the latter directly in a fully discrete manner.

This idea is both general and elementary, derivable in just a few lines of algebra. It can be implemented easily and used in conjunction with any computational tool capable of computing unbroadened Green's functions for any discrete many-body system. In this paper, we illustrate it for quantum impurity models; its transcription to other contexts should be self-evident.  In the DMFT context, our DSR approach allows us to carry out  the self-consistency loop entirely in terms of DSRs of the hybridization function and self-energy, without the need to remove unphysical negative spectral weight.

In \Sec{sec:ModelHamiltonians}, we define the general Anderson impurity model, as well as its local Green's function $G(z)$ and the non-interacting Green's function $G_0(z)$, which define the self-energy we aim to compute.  In \Sec{sec:SigmaFromDiscreteEQM}, we present our method to compute causal and numerically stable self-energies for DSRs, based on Kugler's symmetric improved estimator~\cite{kugler_improved_2022}. \Sec{sec:Applications}  presents numerical examples that demonstrate the performance of the method
for an impurity model and for DMFT\@.

\Appendix{sec:notation} summarizes the notation used for pole expansions and discrete representations throughout the paper.
Depending on the impurity solver employed, correlation functions may be given in different discrete representations.
Based in part on previous literature~\cite{lowdin_non-orthogonality_1950, lowdin_note_1951, lowdin_quantum_1955, andersen_linear_1975, andersen_muffin-tin_2000, gunnarsson_density-functional_1989, anisimov_density-functional_1991, koch_sum_2008,haverkort_multiplet_2012,lu_efficient_2014}, \Appendix{sec:Srepresentations} summarizes several of these representations together with the transformations between them.
In \Appendix{sec:DiscreteG}, we discuss representation-specific algorithms for the calculation of DSR of the self-energy.
In \Appendix{sec:update-impurity-problem}, we show how the impurity problem can be updated using the discrete representation of self-energy $\Sigma(z)$ required for the implementation of the discrete DMFT self-consistency cycle.
Finally, in \Appendix{sec:quasparticle-weight}, we discuss how the calculation of the quasiparticle weight $Z$ can be regularized in the presence of spurious poles at low energies.

\section{\label{sec:ModelHamiltonians}Model Hamiltonian}

The formulation for obtaining the discrete self-energy, as derived in \Eq{eq:Sigma_inv} below, is in principle valid for any Hamiltonian $H = H_0+H_1$, acting on a set of fermionic single-particle states and downfolded onto a subset of these states. For concreteness, and to obtain numerical models that can be solved with the techniques currently available to us, we focus on its application to an Anderson impurity model. In particular, we consider a fermionic multi-orbital quantum impurity model with $N$ local degrees of freedom, labelled by $\mi \in \{1,\hdots,N\}$. The local degrees of freedom may correspond to atomic-like states $\mi = (n,l,m,\sigma)$ or to different cluster sites in extended impurity models. We label the impurity site by $i=0$ and the $M$ discrete bath sites by $i \in \{1,\dots,M\}$. Particles in these states are created and annihilated by the operators $\CR{i,\mi}$ and $\AN{i,\mi}$, respectively. To suppress the spin-orbital index $\mi$ as much as possible, we introduce row-vectors $\CR{i}$ containing the operators  $\CR{i,\mi}$ for all values $N$ of $\mi$,
\begin{align}
    \label{eq:crvecoverm}
    \CR{i} = \left( \CR{i,\mi=1} \, \CR{i,\mi=2} \, \hdots \, \CR{i,\mi=N} \right),
\end{align}
and the column vector $\AN{i} = \left(\CR{i}\right)^{\dag}$.

The Hamiltonian for our model consists of two parts. The term $H_0$ contains contributions quadratic in the creation and annihilation operators and acts non-trivially both on the bath and on the impurity, while $H_1$ is an arbitrary interacting part that only acts non-trivially on the impurity. The full Hamiltonian is given by
\begin{align}
    \label{eq:Himp}
    H = H_0 + H_1 \, ,
\end{align}
with
\begin{align}
    \label{eq:H0imp}
    H_0 & = \sum_{i=0}^{M} \CR{i} E_{i} \AN{i}
    + \sum_{i=1}^M \CR{0} V^{\dag}_{i} \AN{i} + \CR{i} V_{i} \AN{0} \, ,
\end{align}
where $E_{i}$ and $V_{i}$ are $N \times N$ matrices describing the on-site energies (including the chemical potential) and the hybridization between impurity and bath sites, respectively.
While $H_1$ is in principle an arbitrary interacting term which acts non-trivially only on the impurity, it usually exhibits a bilinear and a quartic term,
\begin{align}
    \label{eq:H1imp}
    H_1 & =  \sum_{\mi_1,\mi_2,\mi_3,\mi_4} U_{\mi_1\mi_2\mi_3\mi_4} \CR{0,\mi_1}\CR{0,\mi_2}\AN{0,\mi_3}\AN{0,\mi_4}
    \\ \nonumber
        & -  \sum_{\mi_1,\mi_2} U_{\mi_1\mi_2}^{\mathrm{MF}} \CR{0,\mi_1}\AN{0,\mi_2} \, .
\end{align}
Here, the tensor $U_{\mi_1\mi_2\mi_3\mi_4}$ defines the local interaction on the impurity and $U_{\mi_1\mi_2}^{\mathrm{MF}}$ can be used to subtract the mean-field contributions of the interaction, which in some approaches are already included in $H_0$.

The influence of the bath on impurity dynamics is completely encoded in the hybridization function,
\begin{align}
    \label{eq:DeltaHybDef}
    \Delta(z) = \sum_{i=1}^{M} V^{\dag}_i \frac{1}{z - E_i} V_i \, ,
\end{align}
with $z \in \doubleC$. For real-frequency calculations, we take $z=\omega + \I 0^+$.
There is an inherent freedom in the choice of $E_{i}$ and $V_{i}$ for ${i}\geq 1$,
since any bath representation that results in the same $\Delta(z)$ leads to the same impurity dynamics. From a numerical point of view, it is often advantageous to exploit this freedom. For example, by applying a unitary transformation in the bath subspace, one can obtain a representation in which the bath Hamiltonian is diagonal, i.e.\ with diagonal $E_i$ for $i \geq 1$. Moreover, by ordering the bath sites by energy and grouping degenerate bath states, one can construct a symmetry-preserving~\cite{lowdin_non-orthogonality_1950, lowdin_note_1951, lowdin_quantum_1955, haverkort_multiplet_2012} representation in which $E_i = \epsilon_i I_N$ and $V_i$ is Hermitian. (Here, and in the following we will denote the $N \times N$ identity matrix as $I_N$.) An explicit algorithm for constructing such a representation is given in \Appendix{sec:GLP}.

The calculation of the self-energy of an impurity model via its definition of \Eq{eq:SigmaG0G} requires the knowledge of the single-particle Green's functions $G(z)$ and $G_0(z)$. For an Anderson impurity model, the fermionic impurity Green's function $G(z)$ is defined as an $N \times N$ matrix in spin-orbital space ($\mi$) with elements
\begin{align}
    \label{eq:Gpropgen}
    G(z)_{\mi,\mi'} & = \sum_{s} \rho_s  \Bigg(
    \brakettt{\Psi_{s}}{\AN{0,\mi}\frac{1}{z - H + \mathcal{E}_{s} } \CR{0,\mi'}} {\Psi_{s}}
    \nonumber
    \\
                    & \ + \brakettt{\Psi_{s}}{\CR{0,\mi'}\frac{1}{z + H - \mathcal{E}_{s} } \AN{0,\mi}} {\Psi_{s}} \Bigg) .
\end{align}
Here,
$\ket{\Psi_{s}}$ are the eigenstates of $H$ with eigenenergies $\mathcal{E}_{s}$, and $H = H_0 + H_1$ is the full Hamiltonian and the Boltzmann weight $\rho_s = \frac{e^{-\mathcal{E}_s/(\kB T)}}{ \sum_{s'} e^{-\mathcal{E}_{s'}/(\kB T)}}$, with $T$ the temperature of the system.
$G_0(z)$ is obtained analogously from \Eq{eq:Gpropgen} by replacing $H$ with $H_0$, and $\ket{\Psi_s}$ and $\mathcal{E}_s$ by the eigenstates $\ket{\Psi_{0s}}$ and eigenenergies $\mathcal{E}_{0s}$ of $H_0$. For the Anderson impurity model, it can be written as
\begin{align}
    G_0(z) = \frac{1}{z - E_0 - \Delta(z)},
\end{align}
with $E_0$ and $\Delta(z)$ defined in \Eqs{eq:H0imp} and~\eqref{eq:DeltaHybDef}.

As outlined in the introduction, calculating $\Sigma(z)$ from $G_0(z)$ and $G(z)$ can be numerically challenging. The next section presents a solution that works for discrete representations of Green's functions and guarantees the causality of $\Sigma(z)$ even in the presence of numerical inaccuracies.

\section{\label{sec:SigmaFromDiscreteEQM}Discrete realization of the symmetric improved self-energy estimator}

In Ref.~\cite{kugler_improved_2022}, Kugler showed how the self-energy can be obtained from a combination of two equations of motion. The construction involves an $\tN \times \tN$ matrix-valued propagator $\tG(z)$, defined analogously to \Eq{eq:Gpropgen}, with  $\tN = 2N$. Here, the $N$ impurity operators $\AN{0,\mi}$ appearing in \Eq{eq:Gpropgen} are augmented by $N$ additional auxiliary operators  $q_\mi$, defined as
\begin{align}
    q_\mi = [\AN{0,\mi},H_1] \, .
\end{align}
The propagator
$\tG(z) =
    \begin{smallpmatrix}
        \tG_{11} & \tG_{12} \\
        \tG_{21} & \tG_{22}
    \end{smallpmatrix}
$
therefore has a $2 \times 2$ block structure. Index $1$ refers to the states where one acts with the operators $\AN{0,\mi}$ and $\CR{0,\mi}$ on $\ket{\Psi_s}$, index $2$ refers to the states where one acts with  the operators $q_{\mi}$ and $q^{\dag}_{\mi}$ on $\ket{\Psi_s}$. As a result,  $\tG_{11} (z) = G(z)$
is the conventional impurity propagator defined above.

Kugler showed~\cite{kugler_improved_2022} that the self-energy, defined in \Eq{eq:SigmaG0G}, can be obtained from elements of $\tG(z)$ via~\footnote{
    The notation
    \unexpanded{$\tG_{11} = G$},
    \unexpanded{$\tG_{21} = F^{\mathrm{L}}$},
    \unexpanded{$\tG_{12} = F^{\mathrm{R}}$},
    \unexpanded{$\tG_{22} = I$}
    is used in Ref.~\onlinecite{kugler_improved_2022}
}
\begin{align}
    \label{eq:Kugler_trick}
    \Sigma(z) - \Sigma^\mH = \tG_{22} (z)  - \tG_{21} (z)
    \frac{1}{\tG_{11} (z)}  \tG_{12} (z) \, .
\end{align}
Here, $\Sigma^{\mH}$ is the Hartree--Fock contribution to the self-energy.
It can be calculated with~\cite{kugler_improved_2022}
\begin{align}
    \label{eq:SigmaHdef}
    \left(\Sigma^{\mH}\right)_{\mi,\mi'}
    =
    \langle \{ q_\mi, a_{0,\mi'}^\dagger\} \rangle
    =
    \langle \{ a_{0,\mi}, q_{\mi'}^\dagger\} \rangle \, ,
\end{align}
where $\langle \bullet \rangle$ denotes the thermal expectation value.

Now recall the Schur complement formula,
\begin{align}
    \label{eq:Schur}
    \left( \left[\begin{medpmatrix}
                         \tG_{11} & \tG_{12} \\
                         \tG_{21} & \tG_{22}
                     \end{medpmatrix}^{-1}\right]_{22} \right)^{-1} =
    \tG_{22}  - \tG_{21} \frac{1}{\tG_{11} }  \tG_{12} \, ,
\end{align}
a block matrix identity that holds provided the inverses involved in it exist. (Its name reflects the fact that its right side is known as the Schur complement of $\tG $ w.r.t.\ $\tG_{11}$.) We can thus express the self-energy as
the inverse of the 22-block of the inverse of $\tilde G$:
\begin{align}
    \label{eq:Sigma_inv}
    \Sigma(z) - \Sigma^\mH = \left( \bigl[\tG^{-1}(z) \bigr]_{22} \right)^{-1} \, .
\end{align}
This is the central equation of this paper.
If the impurity solver yields a causal $\tG(z)$, \Eq{eq:Sigma_inv} is guaranteed to result in a causal self-energy. This is because causality of $\tG(z)$ implies causality of $\tG^{-1}(z)$. The subsequent projection to the $22$ sector also preserves causality, and so does the final inversion.

Obtaining a causal $\tG(z)$ is much easier to achieve than highly accurate pole energies and weights. While high accuracy of individual pole energies and weights is usually not guaranteed as soon as approximations are involved, causality is baked into the design of most Hamiltonian-based impurity solvers. Indeed, except for full density-matrix NRG~\cite{Weichselbaum2007,Peters2006}, where minor causality violations can occur, all impurity solvers used in this work are guaranteed to yield a causal $\tG(z)$ and therefore a causal self-energy estimate via \Eq{eq:Sigma_inv}.

The reformulation~\eqref{eq:Sigma_inv} avoids subtracting two nearly canceling quantities,
instead expressing the self-energy directly through a sub-block of the inverse propagator $\tG^{-1}(z)$. It thus offers
a straightforward strategy for computing the self-energy: compute the full $\tG(z)$ as a $2 \! \times \! 2$ block matrix,  then (i) invert it, (ii) project to the 22 block, and (iii) invert again. The three steps (i)-(iii), described in detail in \Appendix{sec:DiscreteG}, can be performed using matrix manipulations not involving $z$.

When inverting $\tG(z)$, it is important to realize that the operators $\AN{0,\mi}$ and $q_{\mi'}$ do not satisfy the canonical fermionic anticommutation relations. Consequently, $\tG(z)$ is not canonically
normalized, in the sense that
\begin{align}\label{eq:DefStilde}
    -\frac{1}{\pi} \int_{-\infty}^{\infty}\D\omega\> \mathrm{Im} \tG(\omega+\I0^+) = \tS \, ,
\end{align}
where the weight matrix $\tS$ is a positive semi-definite Hermitian matrix instead of the identity matrix
with blocks given by
\begin{subequations}
    \begin{align}
        (\tS_{11})_{\mi\mi'} & = \delta_{\mi\mi'} \,      & (\tS_{12})_{\mi\mi'} & = (\Sigma^{\mH})_{\mi\mi'}
        \\
        (\tS_{21})_{\mi\mi'} & = (\Sigma^{\mH})_{\mi\mi'} & (\tS_{22})_{\mi\mi'} & = \langle \{ q^\pdag_{\mi},q^{\dag}_{\mi'}\} \rangle \, .
    \end{align}
\end{subequations}
The block $\tS_{12}$ can be made to vanish by absorbing the Hartree--Fock mean-field part of $H_1$, evaluated from the ground-state one-particle density matrix, into $H_0$, and correspondingly removing it from $H_1$, i.e., setting $U^\mathrm{MF}_{\mi\mi'} = (\Sigma^{\mH})_{\mi\mi'}$ in \Eq{eq:H1imp}.

To invert $\tG(z)$, we express it in terms of an effective energy dependent matrix-valued effective Hamiltonian $\tH_{\mathrm{eff}}(z) \in \doubleC^{\tN \times \tN}$  (see~\cite{lowdin_non-orthogonality_1950,gunnarsson_density-functional_1989,koch_sum_2008,lu_efficient_2014}) as
\begin{align}
    \label{eq:GzResHeff}
    \tG(z) = \tS^{1/2} \frac{1}{z -\tH_{\mathrm{eff}}(z)} \tS^{1/2} \, ,
\end{align}
with $\tS^{1/2}$ the Hermitian principle square root of $\tS$.
The inverse of $\tG(z)$ then becomes
\begin{align}
    \tG^{-1}(z) = \tS^{-1/2} \bigl( {z -\tH_{\mathrm{eff}}(z)} \bigr) \tS^{-1/2} \, ,
\end{align}
and the lower-right block becomes
\begin{align}
    \bigl[ \tG^{-1}(z) \bigr]_{22} & = \bigl[ \tS^{-1} \bigr]_{22} z -
    \bigl[ \tS^{-1/2}  \tH_{\mathrm{eff}}(z) \tS^{-1/2} \bigr]_{22} \, .
\end{align}
Inverting the matrices on both sides yields
\begin{align}
    \label{eq:GzdownfoldedSelfEnergy} \nonumber
     & \left( \bigl[ \tG^{-1}(z) \bigr]_{22} \right)^{-1} = \frac{1}{\bigl[ \tS^{-1} \bigr]_{22} z - \bigl[ \tS^{-1/2}  \tH_{\mathrm{eff}}(z)  \tS^{-1/2} \bigr]_{22}} \\   &\quad\quad\quad= S^{1/2} \frac{1}{ z - S^{1/2} \bigl[ \tS^{-1/2} \tH_{\mathrm{eff}}(z)  \tS^{-1/2} \bigr]_{22} S^{1/2} } S^{1/2} \, .
\end{align}
Here, $S^{1/2}$ is the Hermitian principle square root of
$S = \bigl( [ \tS^{-1}]_{22} \bigr)^{-1}$, which constitutes
the weight matrix of $\Sigma(z) -  \Sigma^{\mH} $.
The resulting list-of-poles representation of the discrete self-energy is
\begin{align}
\label{eq:SE_LOP}
        \Sigma(z) & = \Sigma^\mH + \sum_{i} \frac{\SigmaWeight{i}}{z - \SigmaPole{i}} \, ,
\end{align}
with weights $\SigmaWeight{i}$ and pole positions $\SigmaPole{i}$.
Explicit expressions for $\Sigma(z)$ when $\tG(z)$ is known in various representations are given in the \Appendix{sec:DiscreteG}.

Although shown here for Anderson impurity models, the above construction is more general and applicable to interacting fermionic problems downfolded onto a chosen subspace. This makes it particularly useful for applications that are highly sensitive to causality and moment conservation, such as the extraction of low-energy Fermi-liquid parameters and realization of a fully discrete real-frequency DMFT self-consistency loop, as discussed in the following two benchmark applications.

\section{\label{sec:Applications}Applications}

We now apply our discrete self-energy scheme to two challenging problems, employing various zero-temperature impurity solvers. First, \Sec{sec:overview} gives an overview of the impurity solvers used,
and \Sec{sec:NRG_details} provides details of our benchmark method.
In \Sec{sec:Application1}, we consider the single-impurity Anderson model (SIAM) and compute the low-frequency expansion coefficients of the self-energy.
Then, in \Sec{sec:Application2}, we treat the Hubbard model on the $d \to \infty$ Bethe lattice using discrete DMFT
and compute the $U$-dependent double occupancy, low-order moments, and quasiparticle weight.

\subsection{\label{sec:overview}Overview of impurity solvers}

In the following, we employ four different impurity solvers:
(i) the tangent-space Krylov~(TaSK)~\cite{Kovalska2025,Picoli2026} approach;
(ii)--(iii) natural-impurity-orbital solvers~\cite{lu_efficient_2014,lu_exact_2017,lu_natural-orbital_2019} in two independent implementations, (ii)
\Quanty{}~\cite{QuantyWebpage} and (iii) \RAS{}~\cite{RAS_DMFTWebpage};
and (iv) the numerical renor\-mal\-ization group~(NRG)~\cite{wilson_renormalization_1975,bulla_numerical_2008}.
All four solvers discretize the bath, making them amenable to our DSR of the self-energy.
Below, we show that the DSR yields highly accurate self-energies for all of these solvers, demonstrating the generality and versatility of our scheme.

The TaSK method used in solver (i) is a generic approach to compute spectral functions on top of ground states that are represented as a matrix product state~(MPS)~\cite{Kovalska2025}.
It approximates excitations within the tangent space of the ground state MPS using an otherwise standard Krylov scheme.
In this work, we use TaSK as an impurity solver to determine low-frequency Fermi-liquid coefficients of a SIAM\@.
We therefore employ a logarithmic discretization as in NRG to obtain high resolution at the Fermi level (though TaSK does not rely on energy-scale separation).
To resolve spectra with contributions extending over several decades with a Krylov scheme, TaSK is initialized with a block of states across all energy scales, as explained in Ref.~\cite{Picoli2026}.

The natural-impurity-orbital solver used both in (ii) and (iii) is an exact diagonalization (ED)-based approach that employs a truncated configuration space. The truncation is performed by restricting the occupation of bath orbitals in the natural-impurity-orbital basis~\cite{lu_efficient_2014}, thereby selecting the most relevant configurations~\cite{lu_natural-orbital_2019}.

The NRG approach [solver (iv)] relies on logarithmic discretization and an ensuing energy-scale separation along the Wilson chain~\cite{wilson_renormalization_1975,bulla_numerical_2008}. By iterative diagonalization, it generates a complete but approximate many-body basis~\cite{Anders2005}. Correlation functions can then be computed in textbook fashion~\cite{Weichselbaum2007,Peters2006}.
Our DSR for the self-energy can also be used in NRG, see \Fig{fig:fit_SEdisc}(b).
In all other cases, however,
we choose to employ NRG in the conventional way with a continuous self-energy for benchmarking purposes.
Since this NRG benchmark is used in both Sec.~\ref{sec:Application1} and~\ref{sec:Application2}, we summarize its computational details in the following subsection.

\subsection{\label{sec:NRG_details}Details of NRG benchmark}

Within NRG, quasiparticle parameters such as the quasiparticle weight $Z$ and the scattering-rate coefficient $C$ can be extracted from the
finite-size many-body spectra
obtained from iterative diagonalization using Hewson's renormalized perturbation theory~(RPT)~\cite{Hewson2004,Bauer2007,Nishikawa2010,Kugler2020}.
In short, $Z$ is readily determined from the lowest-energy single-particle excitation. For $C$, the quasiparticle interaction is first deduced from two-particle excitations, and a second-order renormalized expansion of the self-energy in terms of quasiparticle propagators provides the $\omega^2$ coefficient~\cite{hewson_renormalized_2001}.

Applied to a SIAM as in Sec.~\ref{sec:Application1}, this NRG+RPT approach does not involve any dynamical correlation functions and is thus extremely accurate. We also use NRG+RPT to determine the quasiparticle weight for the DMFT solution of the Hubbard model in Sec.~\ref{sec:Application2}. However, in this setting, the hybridization function is determined self-consistently involving NRG continuous self-energies that are obtained via a logarithmic broadening.
Through the DMFT self-consistency the self-energy develops non-trivial features at high energies, which are poorly resolved on a logarithmic grid. Both aspects lead to a small but finite error in the final results.

The NRG results were obtained with the MuNRG package~\cite{lee_adaptive_2016,lee_doublon-holon_2017,lee_computing_2021}, built on top of the QSpace tensor library~\cite{Weichselbaum2012a_QSpace,Weichselbaum2020_QSpace,Weichselbaum2024_QSpace,Weichselbaum2024_QSpaceCode},
allowing us to exploit
SU(2) charge and SU(2) spin symmetry.
For the results in Sec.~\ref{sec:Application1},
we keep up to $4000$ low-energy multiplets during the iterative diagonalization, which provides highly accurate finite-size spectra, and consider different discretization parameters $\Lambda$.
For the results in Sec.~\ref{sec:Application2}, involving continuous self-energies, we used $\Lambda=1.7$ and kept $N_{\mathrm{keep}}=5000$ multiplets during the iterative diagonalization.
We averaged results over $n_z=4$ shifted discretization grids~\cite{Zitko2009} and used an adaptive broadening parameter~\cite{lee_adaptive_2016} $\alpha_z=\alpha/n_z$ with $\alpha=1$.

\subsection{\label{sec:Application1}Low-energy Fermi-liquid behavior in the SIAM}

\begin{figure}[tb]
    \includegraphics[width = \linewidth]{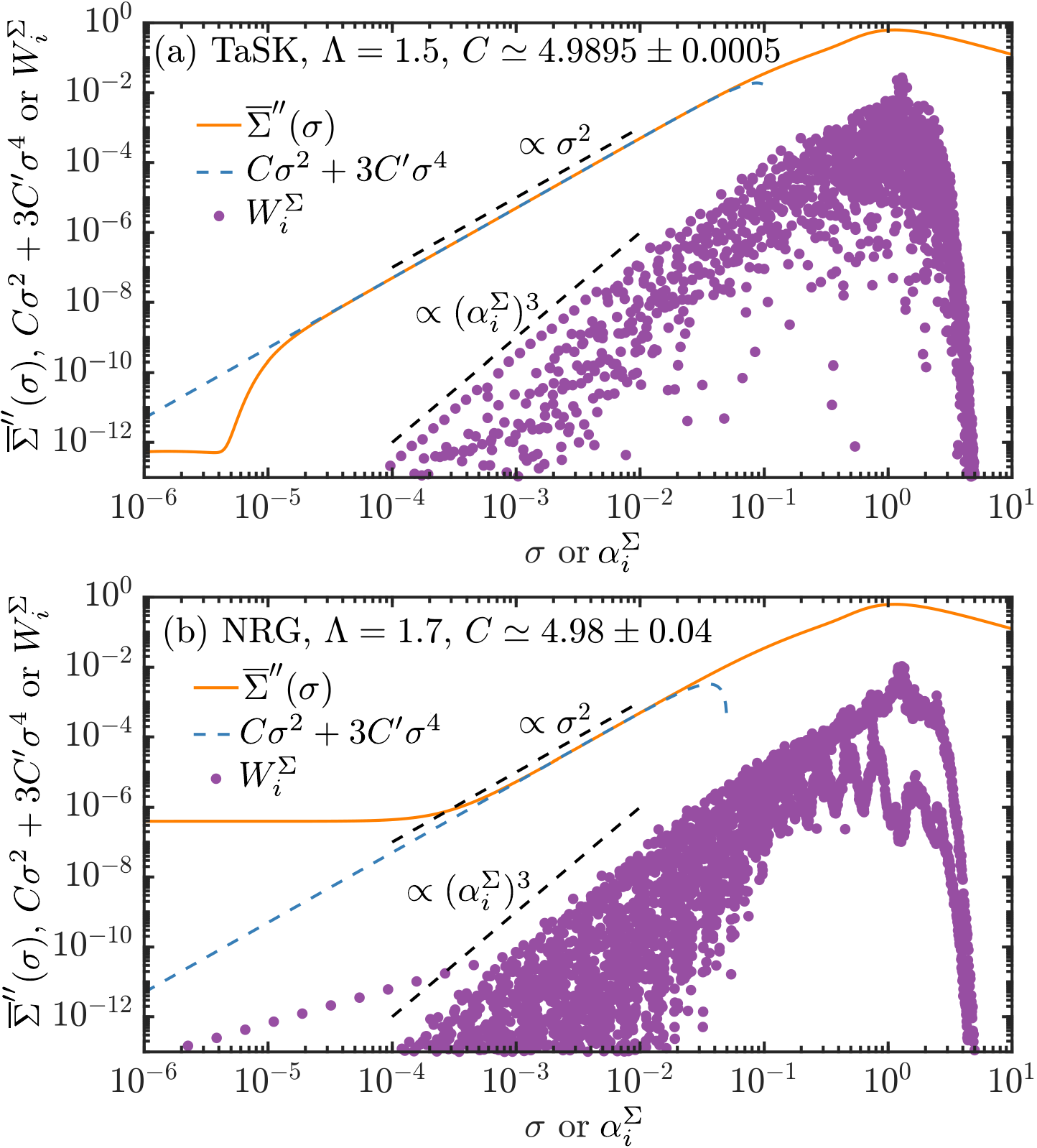}
    \caption{\label{fig:fit_SEdisc}
        Discrete self-energy spectrum $\{\SigmaPole{i}, \SigmaWeight{i}\}$ [c.f.\ Eq.~\eqref{eq:SE_LOP} and \App{sec:notation}] (purple dots) computed for a particle-hole symmetric SIAM with $U=2$ and logarithmically discretized semicircular hybridization density of states [see \Eq{eq:Delta_semicirc}], using (a) TaSK with logarithmic discretization parameter $\Lambda = 1.5$ and (b) NRG with $\Lambda = 1.7$ and plotted only for positive frequencies; the negative-frequency spectrum follows from particle-hole symmetry. The self-energy has the characteristic SIAM structure~\cite{Yamada1975,Hewson1993,Ge2024}: a minimum at $\omega=0$ flanked by two broad peaks positioned symmetrically at $|\omega|
        \gtrsim1$.
        At small discrete frequencies $\SigmaPole{i}$, we find $\SigmaWeight{i} \propto (\SigmaPole{i})^3$. Due to the logarithmic spacing of $\SigmaPole{i}$, the number of weights in an interval $[\omega - \delta\omega,\omega + \delta\omega]$ (with constant $\delta\omega > 0$) around a small frequency $\omega$ is $\propto 1/\omega$. Hence, the total spectral weight in such an interval is $\propto \omega^2$, consistent with Fermi-liquid behavior.
        The orange line shows $\overline{\Sigma}''(\sigma)$ [see \Eq{eq:Ssigma}],
        the Gaussian-weighted average of the discrete spectrum
        with varying kernel width $\sigma$, and the dashed blue line is a fit to $\overline{\Sigma}''(\sigma)$. For panel (a), the fitting range is $\sigma\in[10^{-4},10^{-2}]$, while for panel (b) it is $\sigma\in[10^{-3},10^{-2}]$.
        The coefficient $C$ and its uncertainty are estimated from independent fits of $\overline{\Sigma}''(\sigma)$ over multiple subintervals $\mc{I}_n$ of the corresponding fitting range; the quoted values reflect the fit uncertainty together with the spread among these subinterval fits.
    }
\end{figure}

\begin{figure}[tb]
    \includegraphics[width = \linewidth]{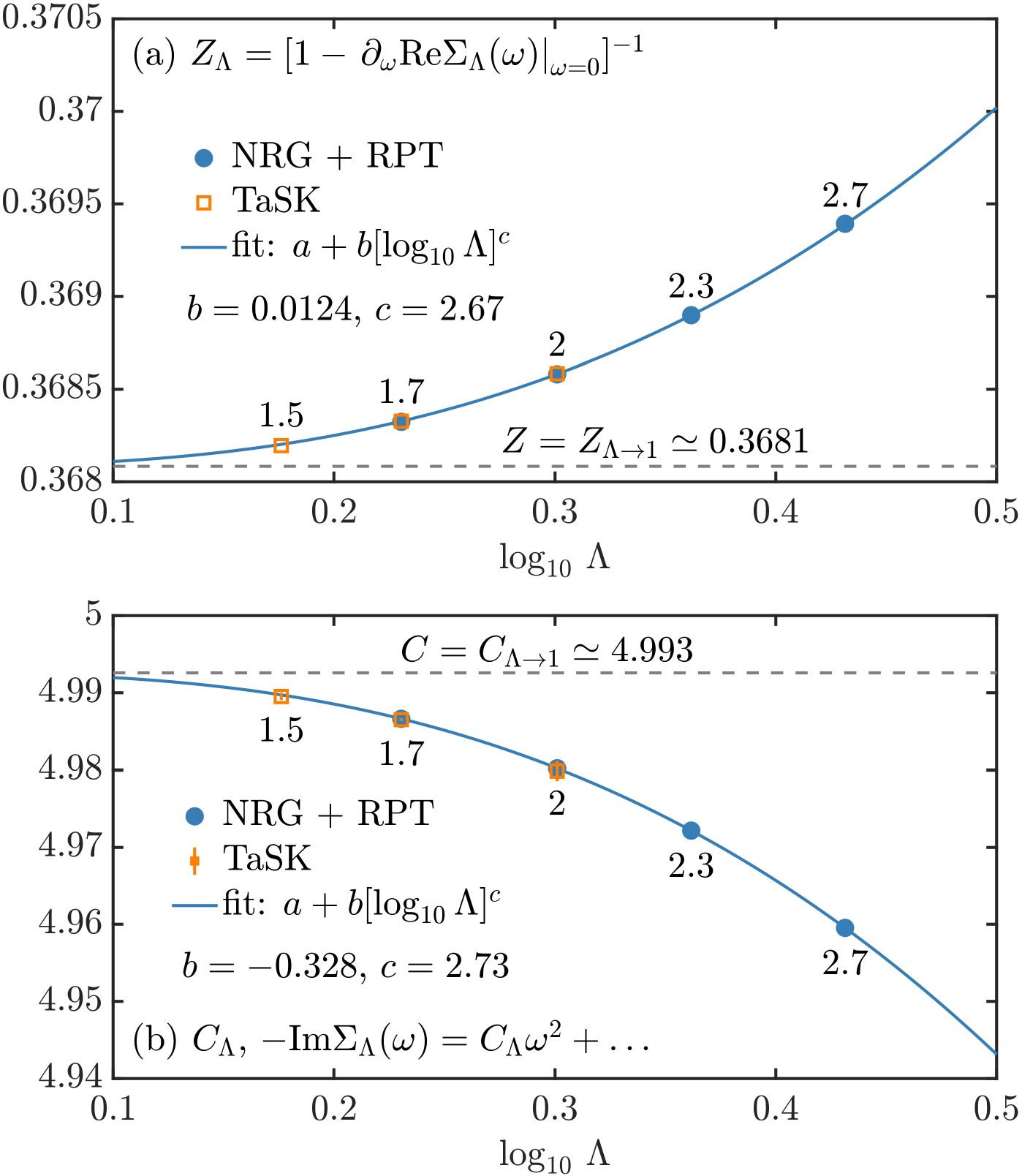}
    \caption{\label{fig:SIAM_lowFrequency}
        (a) Quasiparticle weight $Z$ and (b) $\omega^2$ coefficient $C$ of the spectral part of the self-energy for the same SIAM as in Fig.~\ref{fig:fit_SEdisc}.
        We obtain $Z$ and $C$ either from NRG combined with RPT, or from TaSK via the discrete self-energy spectrum.
        The hybridization function is logarithmically discretized with discretization parameter $\Lambda > 1$ (indicated next to the symbols), and the true continuous hybridization function is recovered in the $\Lambda \to 1$ limit. Both $Z$ and $C$ depend on $\Lambda$; their true values are obtained by extrapolating to $\Lambda = 1$. The solid lines show fits to the NRG+RPT data only, while the TaSK data serves as a consistency check at $\Lambda = 2$ and $\Lambda = 1.7$, and as validation of the fit at $\Lambda = 1.5$.	TaSK does not rely on energy scale separation and can therefore access lower values of $\Lambda$. The error bars of the TaSK results in (b) are estimated from the fitting uncertainty (see text), and are smaller than the symbol size for $\Lambda = 1.7$ and $\Lambda = 1.5$.
    }
\end{figure}

In this section, we apply our method for extracting the discrete spectrum of the self-energy to the SIAM\@.
Our goal is to obtain low-frequency expansion parameters of the self-energy in the limit of a continuous bath, in order to illustrate the ability of our approach to yield high-quality numerical data for the self-energy.

We focus on the particle-hole symmetric, spin-degenerate case (i.e.\ $N=2$), with $E_0 = 0$ and
\begin{align}
    \label{eq:H1SIAM}
    H_1 = U \left(n_{\uparrow} - \tfrac{1}{2}) (n_{\downarrow} - \tfrac{1}{2} \right) \, ,
\end{align}
where $n_{\sigma} = a^{\dag}_{0\sigma} a^{\pdag}_{0\sigma}$ is the spin-$\sigma$ density at the impurity and we set $U = 2$.
Further, we choose a hybridization function with a semicircular density of states with half-bandwidth $D=1$ (our unit of energy),
\begin{align}
    \label{eq:Delta_semicirc}
    -\frac{1}{\pi} \mr{Im} \Delta(\omega) = \Delta''(\omega) = \frac{1}{2\pi}\sqrt{1 - \omega^2} \, .
\end{align}
With decreasing energies, this model flows to a local Fermi-liquid fixed point.

At $T=0$ and given particle-hole symmetry, the impurity self-energy has the low-frequency expansion
\begin{subequations}
    \begin{align}
        \label{eq:SE_lowFrequency}
        \mr{Re} \, \Sigma(\omega^+)  & =  \Sigma(0) + (1 - Z^{-1}) \omega + O(\omega^3)  \, ,
        \\
        -\mr{Im} \, \Sigma(\omega^+) & = C\omega^2 + O(\omega^{4}) \, ,
    \end{align}
\end{subequations}
where $\Sigma(0) = 0$,
consistent with $E_0 = 0$ set above,
and
\begin{align}
    \label{eq:Z}
    Z = [1 - \left. \partial_{\omega} \mr{Re} \, \Sigma(\omega^+) |_{\omega = 0} \right ]^{-1}
\end{align}
is the quasiparticle weight.

Determining the parameters $Z$ and $C$ accurately in the continuum limit is challenging.
A well-established method to access low-energy behavior is to logarithmically  discretize the bath spectrum with a discretization parameter $\Lambda > 1$ and subsequently map the model to a Wilson chain, see Ref.~\cite{bulla_numerical_2008} for details.
The discretized impurity model has discrete spectral properties, from which $\Lambda$-dependent
expansion coefficients $Z_\Lambda$ and $C_\Lambda$ can be extracted with high accuracy. Importantly, this can be done without ever broadening spectral functions, a procedure that introduces additional, somewhat uncontrolled errors.
Finally, $Z_\Lambda$ and $C_\Lambda$ can be extrapolated to the conti\-nuum limit ($\Lambda \to 1$) in which the discretization error vanishes.

Now, NRG is reliable only for $\Lambda \geq 1.7$, since it relies on energy-scale separation.
TaSK, by contrast, does not and hence is reliable also for $\Lambda < 1.7$.
However, TaSK only computes states relevant for specific dynamical correlation functions and does not have access to an entire approximate many-body spectrum, as needed for the current formulation of RPT\@.
Therefore, within TaSK,
$Z$ and $C$ have to be extracted from explicit dynamical calculations of the self-energy; the details of this extraction are explained below.
(That can also be done for NRG self-energies, but the results turn out to be significantly less accurate than those from TaSK.)

Our strategy, therefore, is to compute the $\Lambda$-dependent coefficients $C_{\Lambda}$ and $Z_{\Lambda}$ for $\Lambda \in \{1.7,2,2.3,2.7\}$
using NRG+RPT, and for $\Lambda \in \{1.5,1.7,2\}$ using TaSK with an explicit discrete self-energy calculation.
The TaSK results at $\Lambda = 1.7$ and $\Lambda = 2$ serve as a consistency check against NRG+RPT, while $\Lambda = 1.5$ provides us with a new data point that cannot be accessed reliably with NRG\@. We then extrapolate the $\Lambda$-dependent expansion parameters obtained from NRG+RPT to the continuum limit at $\Lambda = 1$ and use the TaSK data at $\Lambda = 1.5$ for validation. This yields a highly accurate estimate for the true parameters in the presence of a continuous bath.

\textit{Computational details.}
Our TaSK code also uses the QSpace tensor library~\cite{Weichselbaum2012a_QSpace,Weichselbaum2020_QSpace,Weichselbaum2024_QSpace,Weichselbaum2024_QSpaceCode}, allowing us to exploit both SU(2) spin and charge symmetries.
Within TaSK, we first compute the ground state of the discretized impurity model using the density matrix renormalization group~(DMRG)~\cite{White1992,Schollwock2005,Schollwock2011} with controlled bond expansion~(CBE)~\cite{Gleis2023,Gleis2022,Li2024} and mixing~\cite{Hubig2015,Gleis2025}.
To control the bond dimension, we set a singular-value truncation threshold of $\epsilon_{\mr{SVD}} = 10^{-8}$, which results in a numerically exact ground state.
At $\Lambda = 1.5$, this leads to a bond dimension of at most $D^{\ast} = 198$ multiplets.
TaSK then uses the tangent space of this ground state MPS to approximate excited states relevant for dynamical correlation functions~\cite{Kovalska2025}. In contrast to NRG, this tangent space approximation does not discard shell-off-diagonal matrix elements, which is why energy-scale separation is not required for TaSK\@.
As explained in Ref.~\cite{Picoli2026}, TaSK is initialized with a block of initial states across all energy scales, and 20 block Krylov steps are subsequently performed to compute spectral functions.
The resulting discrete self-energy at $\Lambda = 1.5$ is shown in \Fig{fig:fit_SEdisc}(a); \Fig{fig:fit_SEdisc}(b) shows a discrete self-energy spectrum computed with NRG at $\Lambda = 1.7$ for comparison.

\begin{subequations}
    The quasiparticle weight in \Eq{eq:Z} can be readily evaluated for a discrete self-energy,
    \begin{align}
        \Sigma(\omega + \mr{i}0^+) & = \Sigma^\mH + \sum_{i} \frac{\SigmaWeight{i}}{\omega + \mr{i}0^+ - \SigmaPole{i}} \, ,
        \\ \label{eq:dSE}
        \partial_{\omega} \mr{Re} \, \Sigma(\omega + \I0^+) |_{\omega = 0}
                                   & = - \sum_{\SigmaPole{i} \neq 0} \frac{\SigmaWeight{i}}{(\SigmaPole{i})^2} + \sum_{\SigmaPole{i} = 0} \frac{\SigmaWeight{i}}{(0^+)^2} \, .
    \end{align}
\end{subequations}
If $\sum_{\SigmaPole{i} = 0} \SigmaWeight{i} \neq 0$, the second term in \Eq{eq:dSE} diverges, and the quasiparticle weight is zero.
To avoid artifacts from tiny spectral weights at small frequencies in our TaSK calculations, we omit spectral weights where both $\SigmaWeight{i} \leq 10^{-14}$ and $|\SigmaPole{i}| \leq 10^{-14}$.
\Appendix{sec:quasparticle-weight} discusses an alternative regularization scheme.

To obtain the $\omega^2$ coefficient, we
average the self-energy spectrum over a Gaussian kernel of
varying width,
similar to the scheme used in Sec.~IV.C of Ref.~\onlinecite{hanl_equilibrium_2014},
\begin{align}
    \label{eq:Ssigma}
    \overline{\Sigma}''(\sigma) & =  \int_{-\infty}^{\infty} \mr{d}\omega \, \pi\Sigma''(\omega) g_{\sigma}(\omega) \, ,
    \\
    g_\sigma(x)                 & = \frac{1}{\sigma \sqrt{2\pi}} \mr{e}^{-\tfrac{x^2}{2\sigma^2}} \, .
\end{align}
At small frequencies, $\pi\Sigma''(\omega) =  C\omega^2 + C' \omega^4 + \dots$, and, for small width $\sigma \ll 1$, \Eq{eq:Ssigma}  becomes
\begin{align}
    \overline{\Sigma}''(\sigma) = C\sigma^2 + 3 C' \sigma^4 + \dots \, .
\end{align}
From that, we extract $C$ via a log-scale fit.
For this purpose, we choose a fitting interval $\sigma\in \mc{I}$ and perform log-scale fits on multiple subintervals $\mc{I}_n\subset \mc{I}$.
We then discard outlying fits based on the resulting values of $C$ and the fit quality, take $C$ to be the median of the remaining fit values, and estimate its uncertainty by combining the individual fit uncertainties with the spread across subintervals.

Figure~\ref{fig:fit_SEdisc}(a), computed for $\Lambda = 1.5$,
shows TaSK data for the self-energy spectrum (purple dots),
the resulting curve $\overline{\Sigma}''(\sigma)$ (orange), and a fit to it (blue dashed), using the fitting interval $\mc{I} = [10^{-4}, 10^{-2}]$.
Figure~\ref{fig:fit_SEdisc}(b) shows analogous data, obtained for $\Lambda = 1.7$ using NRG\@. We use a smaller fit interval $\mc{I} = [10^{-3}, 10^{-2}]$ here, since the dynamical data provided by NRG is not as clean as that obtained from TaSK\@:
At $|\alpha_i^\Sigma|<10^{-3}$, the NRG data in Fig.~\ref{fig:fit_SEdisc}(b) exhibit spurious poles whose weights scale as $W_i^\Sigma\propto|\alpha_i^\Sigma|$, producing the small-$\sigma$ plateau in $\overline{\Sigma}''(\sigma)$. We suspect that these poles arise from the neglect of inter-shell matrix elements in fdm-NRG; such artifacts are absent in TaSK, which retains inter-shell matrix elements.
For TaSK at $\Lambda = 1.5$, we estimate a relative uncertainty of $0.01 \%$, while the fit to the NRG data has a much larger uncertainty of $0.9\%$.
We have checked that the result obtained from fitting dynamical NRG data differs by $0.14\%$ from the corresponding RPT result, confirming a reasonable albeit possibly somewhat conservative error estimate.
An error of $0.9\%$ is more than an order of magnitude too large for the accuracy requirements of our extraction of $Z_\Lambda$ and $C_\Lambda$ [the error bars would exceed the axis limits of \Fig{fig:SIAM_lowFrequency}(b)]; the dynamical data used for that purpose was therefore based only on TaSK, not on NRG\@.

The $\Lambda$-dependent quasiparticle weight and $\omega^2$ coefficients, fitted either from dynamical TaSK data or obtained through NRG+RPT, are shown in \Fig{fig:SIAM_lowFrequency}(a) and (b), respectively.
For $\Lambda = 1.7$ and $\Lambda = 2$, we performed both NRG+RPT and TaSK computations and find excellent agreement.
We empirically find that both $Z_{\Lambda}$ and $C_{\Lambda}$ follow a power law in terms of $\log \, \Lambda$, which we use to extrapolate to $\Lambda \to 1$ (continuum limit) by fitting the NRG+RPT data. We then validate the extrapolation using the $\Lambda = 1.5$ TaSK result and find excellent agreement: the relative difference between the extrapolated and TaSK results at $\Lambda = 1.5$ is $0.002\%$ for $Z$ and $0.004\%$ for $C$. The resulting extrapolated quasiparticle weight $Z$ and $\omega^2$ coefficient $C$ are highly accurate estimates.

\subsection{\label{sec:Application2}Discrete dynamical mean-field theory}

In this section, we take our scheme a step further and employ the discrete representation of self-energy \Eq{eq:Kugler_trick} to construct a fully discrete DMFT self-consistency cycle, i.e., without introducing any artificial broadening.
Two slightly different implementations of a DMFT(ED) code with restricted Hilbert space,  \Quanty{} and \RAS{}, are employed,
see \Sec{sec:DMFTcomp} for computational details.

We validate this approach on the half-filled Hubbard model on the infinite-dimensional Bethe lattice with nearest-neighbor hopping.
This model has a semicircular density of states,
\begin{align}
    \rho(\epsilon) = \theta(D - |\epsilon|) \frac{2}{\pi D^2} \sqrt{D^2 - \epsilon^2} \, ,
\end{align}
where $D$ is the half bandwidth. We choose $H_0$ to contain only the hopping terms (no chemical potential or on-site energies) and
\begin{align}
    \label{eq:H1DMFT}
    H_1 = U \sum_{i} (n_{i\uparrow} - \tfrac{1}{2})(n_{i\downarrow} - \tfrac{1}{2}) \, .
\end{align}
At half filling and zero temperature, this model exhibits a first-order Mott metal--insulator transition, with metallic and insulating solutions coexisting for $U_{c1} < U < U_{c2}$~\cite{georges_dynamical_1996}.
The available estimates for the boundaries of this coexistence region are roughly $U_{c1}/D \approx 2.4$~\cite{Bulla2001,Garcia2004,karski_single-particle_2008,lee_doublon-holon_2017,Eisenlohr2019} and $U_{c2}/D \approx 2.9$~\cite{Bulla1999,Bulla2001,lee_doublon-holon_2017}.

The effective impurity model that arises when solving this model with DMFT is a SIAM with a self-consistently adjusted hybridization function and an interaction as in \Eq{eq:H1SIAM}.
With this choice of $H_1$,
\begin{subequations}
    \begin{align}
        q_{\sigma}   & = U a_{0\sigma} (n_{\bar\sigma} - \tfrac{1}{2}) \, ,                                      \\
        \tS_{\sigma} & = \diag(1,\langle \{q^{\dag}_{\sigma},q^{\pdag}_{\sigma}\}\rangle) = \diag(1, U^2/4) \, ,
    \end{align}
\end{subequations}
where $\bar{\sigma}$ is the spin projection opposite to $\sigma$ and $n_{\sigma} = a^{\dag}_{0\sigma}a^{\pdag}_{0\sigma}$.

We first confirm that the self-consistency cycle converges to the correct fixed-point solution by comparing the double occupation $\langle n_\uparrow n_\downarrow \rangle$ against reference conventional NRG data with a continuous (broadened) DMFT self-energy. We then test the accuracy of $\Sigma(z)$ directly through the quasiparticle weight $Z$.
We proceed to examine the pole structure of $\Sigma(z)$ in the metallic and insulating phases, where an isolated pole in the insulating gap provides a stringent check via its analytically constrained residue~\cite{georges_dynamical_1996}.
Finally, we verify that the discrete representation satisfies known sum rules for the moments of $\Sigma(z)$, improving on earlier work~\cite{lu_efficient_2014} by several orders of magnitude.

For the implementation of discrete DMFT, the impurity problem needs to be updated using the discrete self-energy $\Sigma(z)$, either via the bath Green's function $\mathcal{G}(z)$ or the hybridization function $\Delta(z)$. In this work, we employ the procedure proposed in Ref.~\cite{lu_efficient_2014}.
The hybridization function is updated as
\begin{align}
    \label{eq:UpdateHybridizationMainText}
    \Delta^{\text{eff}}(z)
    =
    \Delta(z I_N-\Sigma(z))
    =
    \sum_{i=1}^{M} V_i^\dag \frac{1}{z - E_i - \Sigma(z)} V_i \, .
\end{align}
The resulting $\Delta^{\text{eff}}(z)$ is then transformed to the form of \Eq{eq:DeltaHybDef} with a subsequent reduction of the number of poles.
For completeness, we discuss this construction in more detail in \Appendix{sec:update-impurity-problem}.

\subsubsection{Numerical results}

\begin{figure}[tb]
    \includegraphics[scale=1]{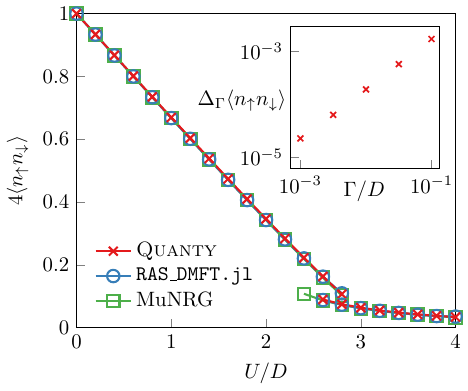}
    \caption{
        \label{fig:double-occupation}
        Double occupation $\langle n_\uparrow n_\downarrow \rangle$ as a function of interaction strength $U/D$ for the half-filled Hubbard model on the Bethe lattice. Results obtained using discrete DMFT with \Quanty{} and \RAS{} are compared to conventional NRG data calculated with MuNRG\@.
        The inset shows the absolute deviation $\Delta_\Gamma \langle n_\uparrow n_\downarrow \rangle = |\langle n_\uparrow n_\downarrow \rangle(\Gamma) - \langle n_\uparrow n_\downarrow \rangle(\text{discrete})|$ at $U/D=2$ as a function of the Lorentzian broadening $\Gamma$ (full width at half maximum) calculated with \Quanty{}.
    }
\end{figure}
We begin by asking whether the discrete self-consistency cycle converges to the same fixed point solution as conventional NRG\@. To test this, in \Fig{fig:double-occupation} we show the double occupation $\langle n_\uparrow n_\downarrow \rangle$ as a function of $U/D$.
The good agreement between \Quanty{}, \RAS{}, and reference NRG data in \Fig{fig:double-occupation} indicates that all approaches converge to similar fixed-point solutions of the DMFT self-consistency cycle. This supports the reliability of the discrete DMFT approach based on a discrete self-energy $\Sigma(z)$ introduced in Sec.~\ref{sec:SigmaFromDiscreteEQM}. We note that with the present parameter set, \Quanty{} and \RAS{} could not converge to the metastable insulating solution at $U/D=2.4$, in contrast to MuNRG\@.

The inset of \Fig{fig:double-occupation} shows the effect of spectral broadening on the double occupation $\langle n_\uparrow n_\downarrow \rangle$.  For $\Gamma>0$, the self-energy is computed using \Eq{eq:Kugler_trick} with $z=\omega + \I\Gamma/2$, corresponding to Lorentzian broadening with full width at half maximum $\Gamma$. The broadening enters only through the calculation of $\Sigma(z)$, and thus affects the DMFT self-consistency, but does not enter the evaluation of $\langle n_\uparrow n_\downarrow \rangle$ directly. As $\Gamma$ decreases, the double occupation converges to the discrete DMFT result, which validates our scheme as a proper $\Gamma\to0$ limit.

\begin{figure}[tb]
    \includegraphics[scale=1]{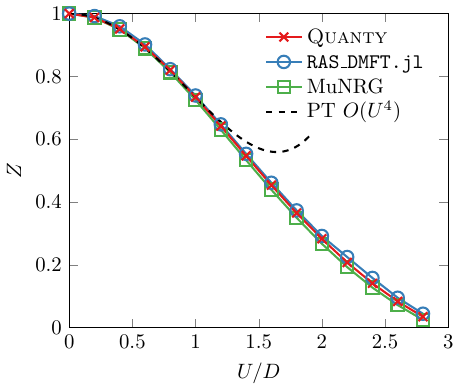}
    \caption{
        \label{fig:quasiparticle-weight}
        Quasiparticle weight $Z$ as a function of interaction strength $U/D$. Results from discrete DMFT (\Quanty{} and \RAS{}) are compared with conventional NRG reference data.
        For small values of $U$, we also compare against fourth-order perturbation theory (PT)
        (Eq.~(113) of Ref.~\cite{Gebhard2003}).
    }
\end{figure}

Next, in \Fig{fig:quasiparticle-weight}, we show the quasiparticle weight $Z$ calculated from \Eq{eq:Z}.
For \Quanty{}  and \RAS{} the quasiparticle weight $Z$ was extracted from the discrete representation of the self-energy using \Eq{eq:dSE}. To deal with spurious poles at low energies, a regularization parameter $\lambda/D=10^{-2}$ was used (See \Appendix{sec:quasparticle-weight} for details). For MuNRG, $Z$ was obtained with NRG+RPT, as described in Sec.~\ref{sec:NRG_details}, after DMFT self-consistency was reached.

A minor discrepancy between the discrete DMFT solvers (\Quanty{} and \RAS{}) and the MuNRG data in \Fig{fig:quasiparticle-weight} may be attributed to finite discretization and truncated Hilbert space used in the former.
Let us emphasize that this is a limitation of the impurity solver, rather than the approach of discrete DMFT or discrete self-energy.
Systematic error introduced by finite broadening~\cite{Zitko_Quantitative} in MuNRG calculations may be another source of discrepancy.
However, note the excellent agreement between MuNRG and fourth-order perturbation theory at low interaction strengths $U$.
An implementation of a similar discrete DMFT scheme in MuNRG is planned for the future.

\begin{figure}[tb]
    \includegraphics[scale=1]{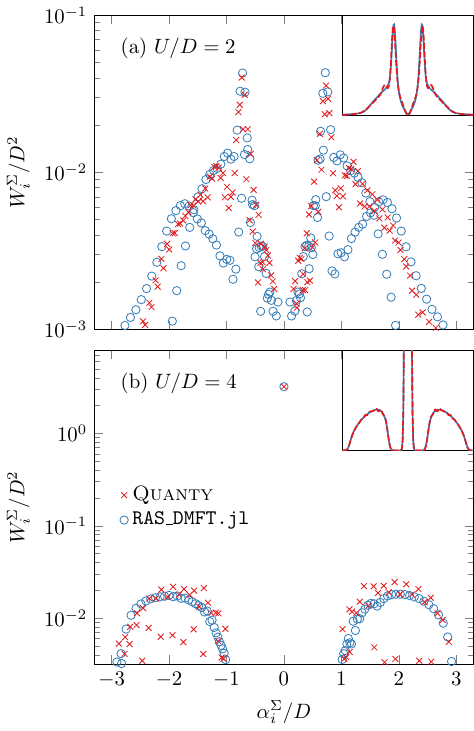}
    \caption{
        \label{fig:self-energy-discrete}
        Discrete self-energy spectrum $\{\SigmaPole{i}, \SigmaWeight{i}\}$ for (a) a metallic solution $(U/D=2)$ and (b) an insulating solution $(U/D=4)$.
        In the metallic case, $\Sigma(\omega)$ exhibits no poles at $\omega=0$,
        keeping the height of the Kondo peak unchanged at zero temperature,
        while in the insulating case, the self-energy exhibits a single pole inside the insulating gap, located at $\omega=0$.
        The residue of this pole is related to the negative second moment of the Green's function [see \Eq{eq:PoleAtw=0}] and shows excellent agreement between the two codes. Particle–hole symmetry was not enforced, demonstrating the numerical stability of the solution. The insets show the respective self-energy convolved with a Gaussian with standard deviation $\sigma/D = 0.08$.
    }
\end{figure}

In \Fig{fig:self-energy-discrete}, we show the discrete residues of the self-energy for the metallic (a) and insulating (b) solutions.
Small discrepancies in the positions or residues of individual self-energy poles between the two codes are not problematic.
Since the total spectral weight $\sum_i \SigmaWeight{i} = S_0^{\Sigma}$ is fixed, the individual residues depend on the number of poles in the discrete representation:
fewer poles necessarily imply larger residues.
Consequently, multiple discrete representations can correspond to the same continuous self-energy once an appropriate broadening is applied.
To illustrate this, the insets show the broadened self-energy, obtained by convolving the discrete poles with a Gaussian.
After broadening, the overall shape of the self-energy $\Sigma(z)$ from the two codes is in very good agreement. Minor discrepancies at high energies, including the local maxima on the outer flanks of the peaks at $\omega\approx\pm1$, visible only in the \Quanty{} result in the inset of \Fig{fig:self-energy-discrete}(a), can be attributed to differences in computational parameters, most importantly the bath discretization ($N_b\approx300$ bath sites in \RAS{} versus $N_b\approx250$ bath sites in \Quanty{}, see \Sec{sec:DMFTcomp}).

For the metallic case [\Fig{fig:self-energy-discrete}(a)], the self-energy is expected to follow a Fermi-liquid behavior in the continuous limit, $-\mathrm{Im}\,\Sigma(\omega) \propto \omega^2$. The discrete residues, however, need not reflect this behavior directly: several nearly degenerate poles may be represented by a single pole with larger residue.
There is, however, one special pole whose residue is constrained: the isolated pole within the insulating gap in \Fig{fig:self-energy-discrete}(b). Its residue $w_0$ satisfies the relation (see Eq.~(237) of Ref.~\cite{georges_dynamical_1996})
\begin{equation}\label{eq:PoleAtw=0}
    w_0^{-1} = -\frac{1}{\pi}\int_{-\infty}^{\infty} \mathrm{d}\omega \frac{\mathrm{Im} G(\omega + \I0^+)}{\omega^2} \, .
\end{equation}
We find excellent agreement for the residue of this pole between the two implementations. Apart from this pole, the self-energy is purely real within the insulating gap.

To demonstrate numerical stability, particle–hole symmetry was not enforced, i.e. $G^{>}(\omega)$ and $G^{<}(\omega)$ were computed independently.
Although the resulting pole structure is only approximately symmetric, these small asymmetries are not significant.
Upon broadening, the self-energy becomes clearly particle–hole symmetric, as evident from the insets of \Fig{fig:self-energy-discrete} and Table~\ref{tab:moment-errors}.

Finally, in Table~\ref{tab:moment-errors}, we compare numerical and analytical moments of the Green's function and the self-energy. We define the $n$-th moment $S_n^{(C)}$ of a generic correlation function $C(z)$ as
\begin{equation}
    \label{eq:moment}
    S_n^{(C)} = -\frac{1}{\pi} \int_{-\infty}^{\infty} \mathrm{d}\omega\>\omega^n \, \mathrm{Im}\, C(\omega+\I0^+).
\end{equation}
For the half-filled Bethe lattice, the first moments of the Green's function and the self-energy are~\cite{potthoff_interpolating_1997}
\begin{subequations}
    \label{eq:self-energy-moments}
    \begin{alignat}{2}
        S_0^{(G)}  & = 1 \, , \qquad & S_2^{(G)}      & = \frac{U^2}{4} + \frac{D^2}{4} \, , \\
        \Sigma^\mH & = 0 \, , \qquad & S_0^{(\Sigma)} & = \frac{U^2}{4} \, ,
    \end{alignat}
\end{subequations}
with every odd moment vanishing due to particle-hole symmetry.
The Hartree--Fock term $\Sigma^\mH$ is zero with $H_1$ defined in \Eq{eq:H1DMFT}.

Compared to Ref.~\cite{lu_efficient_2014}, where the self-energy was calculated using \Eq{eq:SigmaG0G} (instead of \Eq{eq:Sigma_inv}, which is used in this work), we observe an improvement of several orders of magnitude in the accuracy of the zeroth moment of the self-energy $S_0^{(\Sigma)}$, reaching near machine precision. This improvement can be understood from the fact that $S_0^{(\Sigma)}$ is directly given by $\tS$, the zeroth moment of the matrix-valued propagator $\tG(z)$ [see \Eq{eq:DefStilde}]. The norm of self-energy $S_0^{(\Sigma)}$ is therefore expected to be extremely accurate in sum-rule conserving impurity solvers.

\renewcommand{\arraystretch}{1.3}
\begin{table}[t]
    \centering
    \caption{
        \label{tab:moment-errors}
        Moments of the Green's function and self-energy for the half-filled Bethe lattice.
        For non-zero analytical moments we report the relative error
        $\delta S = |S-S_{\mathrm{analytical}}|/S_{\mathrm{analytical}}$, while for moments that vanish due to particle-hole symmetry the numerical value is shown directly.
    }
    \begin{ruledtabular}
        \begin{tabular}{c l d{2.7} d{2.7} d{2.7}}
            \addlinespace
            $U/D$ & \multicolumn{1}{c}{Code} & \multicolumn{1}{c}{$\delta S_0^{(G)}$} & \multicolumn{1}{c}{$\delta S_2^{(G)}$} & \multicolumn{1}{c}{$\delta S_0^{(\Sigma)}$} \\
            \hline
            2     & \Quanty{}                & 2.2\times10^{-16}                      & 2.7\times10^{-7}                       & 6.6\times10^{-14}                           \\
            2     & \RAS{}                   & 1.1\times10^{-15}                      & 1.2\times10^{-8}                       & 1.3\times10^{-14}                           \\
            \addlinespace
            4     & \Quanty{}                & 4.4\times10^{-16}                      & 1.5\times10^{-7}                       & 4.2\times10^{-15}                           \\
            4     & \RAS{}                   & 6.7\times10^{-16}                      & 1.4\times10^{-10}                      & 3.2\times10^{-15}                           \\
            \hline\hline
            \addlinespace
            \multicolumn{5}{c}{Vanishing moments (analytical value $=0$)}                                                                                                    \\
            \hline
            \addlinespace
            $U/D$ & \multicolumn{1}{c}{Code} & \multicolumn{1}{c}{$\Sigma^\mH/D$}     & \multicolumn{1}{c}{$S_1^{(G)}/D$}      & \multicolumn{1}{c}{$S_1^{(\Sigma)}/D^3$}    \\
            \hline
            2     & \Quanty{}                & 2.4\times10^{-7}                       & 2.4\times10^{-7}                       & -9.2\times10^{-7}                           \\
            2     & \RAS{}                   & 5.3\times10^{-8}                       & 5.3\times10^{-8}                       & -1.8\times10^{-7}                           \\
            \addlinespace
            4     & \Quanty{}                & -3.7\times10^{-9}                      & -3.7\times10^{-9}                      & 1.3\times10^{-8}                            \\
            4     & \RAS{}                   & 1.0\times10^{-8}                       & -2.0\times10^{-8}                      & 6.5\times10^{-8}                            \\
        \end{tabular}
    \end{ruledtabular}
\end{table}

\subsubsection{\label{sec:DMFTcomp}Computational Details}

For calculations performed with \RAS{}, the first two conduction and valence bath sites are kept fully unrestricted.
In the remaining conduction and valence chains, up to either a single valence-conduction excitation or up two of valence-impurity or conduction-impurity excitations are allowed.
This corresponds to $L=2$, $p=2$ in notation used in Ref.~\cite{lu_natural-orbital_2019}.
The hybridization function was discretized on a linear grid, resulting in around $N_b=300$ bath sites. The electron addition and removal Green's function $G^{<}(z)$ and $G^{>}(z)$ were calculated independently (i.e.\ particle-hole symmetry was not enforced), with 200 Krylov states calculated at each DMFT iteration.

For calculations using \Quanty{}, two different parameter sets were employed:
For \Figs{fig:double-occupation} and~\ref{fig:quasiparticle-weight}, the bath was discretized on a logarithmic grid, resulting in around $N_b=40$ bath sites.
Only the first sites of valence and conduction chains are kept unrestricted ($L=1$), while allowing for up to two of any: valence-impurity, conduction-impurity, valence-conduction excitations.
This is a larger subspace than $p=2$ used in \RAS{}, since $p=2$ allows only a single valence-conduction excitation.
$G^{>}(z)$ was calculated using 50 states in the Krylov subspace and $G^{<}(z)$ was set equal to $G^{>}(z)$, enforcing particle-hole symmetry.

For \Fig{fig:self-energy-discrete} and Table~\ref{tab:moment-errors}, the bath was discretized on a linear grid, resulting in around $N_b=250$ bath sites.
The configuration space was reduced, allowing for only one of the valence-impurity, conduction-impurity or valence-conduction excitations.
$G^{>}(z)$ and $G^{<}(z)$ were computed independently, with 150 states in the Krylov subspace.
This set of parameters provides improved high-energy resolution and greater number of poles in the self-energy, which is beneficial for visualization in \Fig{fig:self-energy-discrete}.

The data for $\Gamma>0$ shown in the inset of \Fig{fig:double-occupation} were obtained by broadening the four components of
$\tG(z)$ with a Lorentzian kernel on a linear frequency grid of 50001 points. The continuous self-energy is calculated via \Eq{eq:Kugler_trick} and discretized afterward. The hybridization function is then updated via \Eq{eq:UpdateHybridizationMainText}.

\section{Conclusion}

We showed that Kugler's symmetric self-energy estimator~\cite{kugler_improved_2022} $\Sigma(z)$, can be computed by downfolding an augmented matrix-valued propagator $\tG(z)$, i.e., by inverting, projecting, and re-inverting it. These operations can be carried out using numerically stable matrix manipulations on a discrete spectral representation of $\tG(z)$, yielding a discrete spectral representation of
$\Sigma(z)$. This construction avoids the systematic errors that arise in direct Dyson subtraction. In contrast to the Dyson equation, whose formal correctness does not prevent numerical loss of accuracy in floating-point arithmetic, the present method preserves causality and proper normalization also at the numerical level.
Further, Kugler's symmetric self-energy estimator can be extended to irreducible cumulants of Hubbard operators, as explained in Sec.~IX A of the Supplemental Material of Ref.~\cite{Gleis2026}. Our results also apply to this case.

We discussed two prominent zero-temperature applications of the discrete self-energy: the determination of low-energy properties of the single-impurity Anderson model and the implementation of the DMFT self-consistency cycle using only DSRs. In the former case, we accurately determine the $\omega^2$ coefficient $C$ of
$\mathrm{Im}\Sigma(\omega^+)$
and the quasiparticle weight $Z$ for the continuous-bath limit. In the latter, we demonstrate the reliability of the discrete DMFT approach by computing the double occupancy $\langle n_\uparrow n_\downarrow \rangle$, the quasiparticle weight $Z$, and the first moments of the self-energy on the $d\to\infty$ Bethe lattice, all of which are in very good agreement with reference NRG data and analytical results.

\section{Acknowledgements}

We thank Martin Braß and Markus Wallerberger for valuable discussions. We acknowledge funding via the Research Unit ‘QUAST’
by the Deutsche Forschungsgemeinschaft (DFG;\ Project ID FOR5249, Project No.~449872909)
and the Austrian Science Fund (FWF, Project DOI 10.55776/KIN2563725).
FBK acknowledges funding from the \foreignlanguage{german}{Ministerium f\"ur Kultur und Wissenschaft des Landes Nordrhein-Westfalen (NRW-R\"uckkehrprogramm)}; FTE and KH by the FWF project DOI 10.55776/P36213.
AG is supported by the Abrahams Postdoctoral Fellowship of the Center for Materials Theory at Rutgers University.
The work of MP and JvD was supported in part by the Deutsche Forschungsgemeinschaft under Germany's Excellence Strategy EXC-2111 (Project No.~390814868). This work is part of the  Munich Quantum Valley, supported by the Bavarian state government with funds from the \foreignlanguage{german}{Hightech Agenda Bayern Plus}.

\section*{Data availability}

The data that support the findings of this article are openly available~\cite{zacinskis_2026_22829586}.

\appendix

\section{\label{sec:notation} Notation}

Throughout this work, response functions (e.g. Green's functions $G(z)$, hybridization functions $\Delta(z)$, self-energies $\Sigma(z)$, and effective Hamiltonians $H_{\mathrm{eff}}(z)$) are matrix-valued functions of the complex variable $z$.
We write
matrix inverses and Hermitian principal square roots as matrix powers, e.g.\ $G^{-1}(z)$ and
$S^{1/2}$, but resolvents as fractions, e.g.\ $1/(z-H)$.

Response functions can be represented discretely as a list of poles [\Eq{eq:LPdef}], or as a subspace of the resolvent of a larger auxiliary matrix. In the latter case, the block structure chosen for the auxiliary matrix distinguishes the block-tridiagonal (chain) [\Eq{eq:Tridef0}] from the Anderson (star) [\Eq{eq:ANDdef}] representation.

For response functions in the list-of-poles representation [\Eq{eq:LPdef}], we use $\GPole{i}$ for the pole energies, $\GWeight{i}$ for their matrix-valued weights, and $\GHyb{i}$ for the coupling strengths, with $\GWeight{i}=\GHyb{i}^{\dag}\GHyb{i}=\GHyb{i}\GHyb{i}$ (see \Appendix{sec:symeigrepLP}). To show which response function a symbol belongs to, we decorate these three letters: a tilde for the augmented propagator $\tG(z)$, a bar for the effective Hamiltonian $\tH_{\mathrm{eff}}(z)$, and a superscript $\Sigma$ for the self-energy $\Sigma(z)$. The hybridization function $\Delta(z)$ is an exception, where we keep the usual $E_i$ and $V_i$. Table~\ref{tab:PoleExpansionNotation} lists the resulting symbols for each response function.
\begin{table}[h!]
\centering
{
\caption{Notation for pole energies, coupling strengths and weights for the response functions used in this work, listed in the order they appear in the text.}
\label{tab:PoleExpansionNotation}
\begin{tabular}{r|c|c|c}

&  Energy & Coupling &  Weight \\
\hline
Hybridization function $\Delta(z)$ & $E_i$ & $V_i $ &   \\
Self-energy $\Sigma(z)$ & $\SigmaPole{i}$ & $\SigmaHyb{i}$ & $ \SigmaWeight{i}$  \\
Generic propagator $G(z)$ & $\GPole{i}$ & $\GHyb{i}$ & $\GWeight{i}$  \\
Augmented propagator $\tG(z)$ & $\AugGPole{i} $ & $\AugGHyb{i} $  & $\AugGWeight{i} $  \\
Effective Hamiltonian $\tH_\mathrm{eff}(z)$ & $\HeffPole{i}$  & $\HeffHyb{i}$ & $\HeffWeight{i}$  \\
Effective Hybridization $\Delta^\mathrm{eff}(z)$ & $E^{\mathrm{eff}}_i$ & $V^{\mathrm{eff}}_i $ &   \\
\end{tabular}
}
\end{table}

The tridiagonal [\Eq{eq:TRIdef}] and Anderson [\Eq{eq:GAnd}] representations, introduced in \Appendix{sec:Srepresentations}, both involve submatrices $A_i$ and $B_i$. We distinguish these by font: calligraphic $\TRI{A}_i$, $\TRI{B}_i$ in the tridiagonal case, and typewriter $\AND{A}_i$, $\AND{B}_i$ in the Anderson case.

In the following we will write formulas in terms of the generic propagator $G(z)$.
We explicitly include a constant term $A_0=\TRI{A}_0=\AND{A}_0$, which is zero for physical Green's functions, but may be non-zero for the self-energy $\Sigma(z)$.

\section{\label{sec:Srepresentations} Discrete representations of response functions}
When manipulating response functions (e.g.\ Green's function $G(z)$, hybridization function $\Delta(z)$, or self-energy $\Sigma(z)$) it is often convenient to switch between different discrete representations. In this appendix we summarize the relevant transformations as implemented in \Quanty{} and \RAS{} below, following the work of L{\"o}wdin and others~\cite{lowdin_non-orthogonality_1950, lowdin_note_1951, lowdin_quantum_1955, andersen_linear_1975, andersen_muffin-tin_2000, gunnarsson_density-functional_1989, anisimov_density-functional_1991, koch_sum_2008, haverkort_multiplet_2012,lu_efficient_2014}.

There are several ways to represent these functions on a discrete grid. One common method, which is also used throughout the manuscript, is the pole expansion, which we refer to as the list-of-poles representation (\Appendix{sec:GLP}):
\begin{align}\label{eq:LPdef}
    G(z) = A_0 + \sum_{i=1}^{M} \GHyb{i} \frac{1}{z - \GPole{i}} \GHyb{i} \, ,
\end{align}

Besides a pole expansion, other possible discrete representations of response function include a tridiagonal (continued fraction or chain) representation (\Appendix{sec:GTri})
\begin{align}\label{eq:Tridef0}
    G(z) & = \TRI{A}_0 + \TRI{B}_0
    \frac{1}{z - \TRI{A}_1 - \TRI{B}_1
        \frac{1}{z - \TRI{A}_2 - \hdots}
        \TRI{B}_1}
    \TRI{B}_0 \, ,
\end{align}
Anderson (star) representation (\Appendix{sec:GAnd})
\begin{align}\label{eq:ANDdef}
    G(z) = \AND{A}_0 + \AND{B}_0
    \frac{1}{z - \AND{A}_1  - \sum_{i=1}^{M'} \AND{B}_{i}
        \frac{1}{z - \AND{a}_{i+1}}\, \AND{B}_{i}}
    \, \AND{B}_0 \, ,
\end{align}
as well as a mix of chain and star representations, such as the natural-impurity-orbital representation~\cite{lu_efficient_2014}, which is central to the natural-orbital impurity solver~\cite{lu_natural-orbital_2019}.

\subsection{\label{sec:GLP}List-of-poles representation}

We can represent response functions on a discrete set of energies $\GPole{i} \in \doubleR$, with $i\in\{1,\ldots,M\}$, which may be viewed both as pole positions and as points of an energy mesh:
\begin{align}
    \label{eq:GLPdef}
    G(z) = A_0 + \sum_{i=1}^{M} \frac{\GWeight{i}}{z - \GPole{i}} \, ,
\end{align}
with positive semidefinite, Hermitian spectral weight matrices $\GWeight{i}$. Those may be decomposed as
\begin{align}
    \label{eq:GLPdefWi}
    \GWeight{i} = \GHyb{i} \GHyb{i} \, ,
\end{align}
where we choose $\GHyb{i}$ to be the Hermitian principal square root of $\GWeight{i}$, i.e., $\GHyb{i} = \GHyb{i}^\dagger$. This yields a unique factorization that preserves, when present, the symmetry-imposed block structure of $\GWeight{i}$. Moreover, storing and manipulating $\GHyb{i}$ instead of $\GWeight{i}$ ensures that $\GWeight{i} = \GHyb{i}^{\dag} \GHyb{i}$ remains Hermitian and positive semidefinite by construction, even in the presence of numerical noise.

\subsubsection{\label{sec:symeigrepLP}Symmetric eigenbasis representation}

Throughout this work, we represent pole expansions of response functions in a symmetric form, i.e.\ with Hermitian couplings $B_i = B_i^{\dagger}$, and in the eigenbasis of each pole term, i.e.\ such that the pole energy is the same for all spin-orbitals within a given term of the pole expansion. In other words, we use \Eq{eq:GLPdef} rather than the seemingly more general form
\begin{align}
    \label{eq:GLPdefGen}
    G(z) = A'_0 + \sum_{j=1}^{M'} {B'}_j^\dag \frac{1}{z - A'_j} {B'_j} \, ,
\end{align}
where $A'_j$ is a general Hermitian matrix and $B'_j$ is a general complex matrix. The pole expansion in \Eq{eq:GLPdefGen} can always be rewritten in the more symmetric form of \Eq{eq:GLPdef}, which is advantageous for numerical stability with guaranteed causality and for the construction of discretization.

To transform \Eq{eq:GLPdefGen} into the symmetric eigenbasis representation of \Eq{eq:GLPdef}, we first diagonalize the matrix $A'_j$. Let $U_j$ contain the eigenvectors of $A'_j$ in its columns, and let $\alpha_{j\mi}$ denote the corresponding $\mi$-th eigenvalue, such that
\begin{align}
    U_j^\dag A'_j U_j = \diag(\alpha_{j\mi=1}, \hdots, \alpha_{j\mi=N}) \, .
\end{align}

We then insert the identity, $U_j^\pdag U_j^\dag$ twice, into \Eq{eq:GLPdefGen}, which yields
\begin{align}
    G(z) & = A'_0 + \sum_{j=1}^{M'} {B'}_j^{\dag} \frac{1}{z - U_j^\pdag U_j^\dag A'_j U_j^\pdag U_j^\dag} {B'_j} \nonumber                         \\
         & = A'_0 + \sum_{j=1}^{M'} {B'}_j^{\dag} \frac{1}{U_j^\pdag \left(z - \diag(\alpha_{j1}, \hdots, \alpha_{jN})\right) U_j^\dag} {B'_j} \, .
\end{align}
We now use that $(U (z-H) U^\dag)^{-1} = \left({U^\dag}\right)^{-1} (z-H)^{-1} \times$ $U^{-1}$ $= U (z-H)^{-1} U^\dag$ and find
\begin{align}
    G(z) & = A'_0 + \sum_{j=1}^{M'} {B'}_j^{\dag} U_j^\pdag \frac{1}{z - \diag(\alpha_{j1}, \hdots, \alpha_{jN})} U_j^\dag {B'_j} \, .
\end{align}

We can split the single term for each $j$ into $N$ terms, corresponding to poles at the eigenvalues $\alpha_{j\mi}$. The spectral weight at the pole $z=\alpha_{j\mi}$ is then
the outer product of the row vectors
\begin{align}
    W_{j\mi} = \left(\left(U_j^\dag B'_j\right)_{\mi}\right)^\dag
    \left(\left(U_j^\dag B'_j \right)_{\mi}\right) \, ,
\end{align}
and the corresponding coupling matrix is
\begin{align}
    B_{j\mi} = W_{j\mi}^{1/2} \, .
\end{align}
Thus, the symmetric pole expansion of \(G(z)\) in its eigenbasis becomes
\begin{align}
    G(z) = A'_0 + \sum_{j=1}^{M'} \sum_{\mi=1}^{N} B_{j\mi} \frac{1}{z-\alpha_{j\mi}} B_{j\mi} \, .
\end{align}
Finally, one may combine the sums over $j$ and $\mi$ into a single sum over $i$, with $i=1,\ldots,M=M'N$, to recover the form of \Eq{eq:GLPdef}. Note that one can combine two poles at the same energy $\alpha_i=\alpha_j$ by summing their spectral weights $W_i$ and $W_j$. A general method to reduce the number of poles can be found in Appendix F of Ref.~\cite{lu_efficient_2014}.

\subsection{\label{sec:GTri}Tridiagonal (chain-impurity) representation}

Instead of representing a discretized Green's function as a sum over poles and residues, we may represent it as the Green's function of a local site coupled to a one-dimensional chain of bath sites. This yields a matrix-valued continued-fraction representation,
\begin{align}\label{eq:TRIdef}
    G(z) & = \TRI{A}_0 \nonumber \\
         & \quad + \TRI{B}_0
    \frac{1}{z - \TRI{A}_1 - \TRI{B}_1
        \frac{1}{z - \TRI{A}_2 - \hdots}
        \TRI{B}_1}
    \TRI{B}_0 \, ,
\end{align}
where $\TRI{A}_i$, $\TRI{B}_i \in \doubleC^{N \times N}$ are Hermitian matrices, and the chain has length $M$.

Equivalently, $G(z)$ can be written as the upper-left $N\times N$ block of the resolvent of a block-tridiagonal Hamiltonian $\Ht \in \doubleC^{NM \times NM}$,
\begin{align}
    G(z) = \TRI{A}_0 + \TRI{B}_0 \left(\frac{1}{z - \Ht}\right)_{1:N,\,1:N} \TRI{B}_0 \, ,
\end{align}
with
\begin{align}
    \label{eq:Htridiagonal}
    \Ht =
    \begin{pmatrix}
        \TRI{A}_1 & \TRI{B}_1 & 0      & \cdots        & 0             \\
        \TRI{B}_1 & \TRI{A}_2 & \ddots & \ddots        & \vdots        \\
        0         & \ddots    & \ddots & \ddots        & 0             \\
        \vdots    & \ddots    & \ddots & \TRI{A}_{M-1} & \TRI{B}_{M-1} \\
        0         & \cdots    & 0      & \TRI{B}_{M-1} & \TRI{A}_M
    \end{pmatrix}.
\end{align}

This representation has a natural interpretation as a non-interacting tight-binding chain. Writing the corresponding second-quantized Hamiltonian as
\begin{align}
    \Ht & = \sum_{i=1}^{M}  \CR{i} \TRI{A}_i \AN{i} + \sum_{i=1}^{M-1} \CR{i}
    \TRI{B}_i \AN{i+1} + \CR{i+1}
    \TRI{B}_i \AN{i} \, ,
\end{align}
using the compact vector notation from \Eq{eq:crvecoverm} to suppress the summation over the local spin-orbital indices $\mi$,
we can express $G(z)$ as the local Green's function at the endpoint $(i=1)$ of the one-dimensional chain,
\begin{align}
    G(z)_{\mi,\mi'} =
    \brakettt{0}{\AN{\mi,i=1}\,\frac{1}{z - \Ht}\,\CR{\mi',i=1}}{0} \, .
\end{align}
The tridiagonal representation of $G(z)$ therefore maps the original problem onto a non-interacting one-dimensional tight-binding chain of length $M$, with the probed site located at one end.

\paragraph{\label{sec:LP->Tri}Construction from the list-of-poles representation.}

The matrices $\TRI{A}_i$ and $\TRI{B}_i$ can be obtained from the list-of-poles representation.
For the constant term, we have
\begin{align}
    \TRI{A}_0 = A_0 \, .
\end{align}

In order to find the other block matrices we define the block-vector $\BlockVector \in \doubleC^{NM \times N}$ as
\begin{equation}
    \BlockVector =
    \begin{pmatrix}
        \GHyb{1}    \\
        \GHyb{2}    \\
        \vdots \\
        \GHyb{M}
    \end{pmatrix},
\end{equation}
where the $\GHyb{i} \in \doubleC^{N \times N}$ are the principal square roots of the residues in \Eq{eq:GLPdef}. As the column vectors of $\BlockVector$ are not orthonormal for general response functions, we define the overlap matrix $S \in \doubleC^{N \times N}$

\begin{equation}
    S
    =
    \BlockVector^\dag \BlockVector
    =
    \sum_{i=1}^M \GHyb{i} \GHyb{i}
    =
    \sum_{i=1}^M \GWeight{i} \, ,
\end{equation}
with $\GWeight{i}$ defined in \Eq{eq:GLPdefWi}.
This allows us to write
\begin{align}
    R_a       & = \BlockVector S^{-1/2} \, , \nonumber \\
    \TRI{B}_0 & = S^{1/2} \, ,
\end{align}
with $R_a^\dag R_a = I_N$.

We further define the diagonal matrix
\begin{align}
    H_{\mathrm{LP}} = \diag\!\left(\GPole{1} I_N,\, \GPole{2} I_N,\, \ldots,\, \GPole{M} I_N\right) \in \doubleR^{NM \times NM} \, ,
\end{align}
where $\GPole{i}$ are the pole energies from \Eq{eq:GLPdef}.

Applying a block-Lanczos tridiagonalization to $H_{\mathrm{LP}}$ with starting block $\KrylovState_0 = R_a$ yields the block-tridiagonal matrix $\Ht$ of \Eq{eq:Htridiagonal}, from which the matrices $\TRI{A}_i$ and $\TRI{B}_i$ can be read off. Standard block- or band-Lanczos implementations often produce a banded form with minimal bandwidth, in which $\TRI{B}_i$ is block upper or lower triangular and therefore not Hermitian. It is important to keep $\TRI{B}_i$ Hermitian in order to preserve symmetric couplings between the bath sites.
In the next paragraph we discuss L{\"o}wdin symmetric tridiagonalization, which can be used to ensure Hermiticity of $\TRI{B}_i$.
\paragraph{\label{sec:symmetricLanczos}L{\"o}wdin symmetric tridiagonalization using block Lanczos.}

We can determine $\Ht$ using a block-Lanczos algorithm.
Given an effective Hamiltonian $H_\mathrm{eff} \in \doubleC^{NM \times NM}$
and an orthonormal block vector $\KrylovState_i \in \doubleC^{NM \times N}$,
whose columns contain the $N$ orthonormal Krylov states of block $i$, we construct the Hermitian matrices $\TRI{A}_i$, $\TRI{B}_{i} \in \doubleC^{N \times N}$, as well as the next set of Krylov states $\KrylovState_{i+1} \in \doubleC^{NM \times N}$, as follows.

We first define the block vector
\begin{align}
    \overline{\KrylovState}_{i+1} = H_\mathrm{eff} \KrylovState_i \, .
\end{align}
From \(\overline{\KrylovState}_{i+1}\), we obtain \(\TRI{A}_i\) as
\begin{align}
    \TRI{A}_i = \KrylovState_i^\dag \overline{\KrylovState}_{i+1}^\pdag \, .
\end{align}
Next, we orthogonalize the $N$ vectors in $\overline{\KrylovState}_{i+1}$ against all previously determined Krylov states,
\begin{equation}
    \overline{\overline{\KrylovState}}_{i+1}
    =
    \overline{\KrylovState}_{i+1}
    - \KrylovState_i \bigl(\KrylovState_i^\dag \, \overline{\KrylovState}_{i+1}^\pdag\bigr)
    - \KrylovState_{i-1} \bigl(\KrylovState_{i-1}^\dag \, \overline{\KrylovState}_{i+1}^\pdag\bigr) \, .
\end{equation}
The orthogonalization against $\KrylovState_{i-2}$, $\KrylovState_{i-3}$, and earlier states is not necessary,
as $\overline{\KrylovState}_{i+1}$ is already orthogonal against those states by construction.

At this point the $N$ column vectors inside $\overline{\overline{\KrylovState}}_{i+1}$ are orthogonal to all column vectors inside $\KrylovState_j$ of blocks $j \in \{1,2,\ldots,i\}$: $\KrylovState_j^\dag \KrylovState_{i+1} = 0_N$, but not necessarily to each other: $\overline{\overline{\KrylovState}}_{i+1}^\dag \overline{\overline{\KrylovState}}_{i+1} \neq I_N$.

We now need to orthonormalize $\overline{\overline{\KrylovState}}_{i+1}$ symmetrically. To this end, we define the overlap matrix.
\begin{align}
    \overline{\overline{S}}_{i+1}
    = \overline{\overline{\KrylovState}}_{i+1}^\dag\,\overline{\overline{\KrylovState}}_{i+1} \, ,
\end{align}
and obtain the orthonormal Krylov states as
\begin{align}
    \label{eq:orhtostepinblockLan}
    \KrylovState_{i+1} = \overline{\overline{\KrylovState}}_{i+1}\,\overline{\overline{S}}_{i+1}^{-1/2}.
\end{align}
Finally, we compute $\TRI{B}_i$ as
\begin{equation}
    \TRI{B}_i = \overline{\overline{S}}_{i+1}^{1/2} \, .
\end{equation}

Note that \Eq{eq:orhtostepinblockLan} requires the inverse square root of the overlap matrix. This is not possible if $\overline{\overline{S}}_{i+1}$ is singular. In that case, some form of deflation~\cite{baiSymmetricBandLanczos2001} must be implemented. One may obtain $\KrylovState_{i+1}$ from $\overline{\overline{\KrylovState}}_{i+1}$ by performing a singular value decomposition of the column matrix $\overline{\overline{\KrylovState}}_{i+1}$ and discarding states whose singular values fall below a chosen numerical threshold.
\subsection{\label{sec:GAnd}Anderson (star-impurity) representation}

In the Anderson (or star) representation, the bath degrees of freedom are diagonalized. The Green's function then takes the form
\begin{align}
    \label{eq:GAnd}
    G(z) = \AND{A}_0 + \AND{B}_0
    \frac{1}{z - \AND{A}_1  - \sum_{i=1}^{M'} \AND{B}_{i}
        \frac{1}{z - \AND{a}_{i+1}}\, \AND{B}_{i}}
    \, \AND{B}_0 \, ,
\end{align}
with the impurity at site index $1$ and the bath sites at indices $2$ to $M'+1$.

Note that the number of bath sites and their internal dimensions are, to some extent, a matter of representation. We allow bath sites that do not couple to the impurity in order to keep the internal dimension of each bath site equal to $N$. This increases the number of bath sites to at most $M' = N(M-1)$ for a $N \times N$ matrix valued Green's function with $M$ poles. Keeping the bath dimension fixed can be useful in multi-spin-orbital situations with symmetry. By representing $\AND{B}_i$ as a Hermitian matrix, different spin-orbit states that belong to the same multidimensional irreducible representation have no coupling between them. This is important to conserve numerically the symmetry of the system. At the same time spin-orbitals belonging to different irreducible representations may couple to bath states at different energies, thereby making the rank of $\AND{B}_i$ smaller than $N$. If the numerical storage of the additional zeros is a concern one can transform the spin-orbital basis of the Green's function to symmetric states, e.g., using a spherical tensor coupling and only store the irreducible elements~\cite{tagliavini_polarization_2025}. The symmetric representation on the eigenbasis of the bath is also useful when one wants to reduce the number of bath sites. Since the denominator in the Anderson representation is itself a Green's function represented as a list of poles, one may reduce the number of poles by discretizing this function on a newly chosen grid, as described in Appendix F of Ref.~\cite{lu_efficient_2014}.

We choose the hybridization between the impurity and the bath level at energy $\AND{a}_{i+1}$ to be described by a Hermitian matrix $\AND{B}_i$. Since each bath site of dimension $N$ has all spin-orbitals at the same energy, we may apply a unitary rotation within that bath site, which allows us to choose $\AND{B}_i$ Hermitian. See \Appendix{sec:symeigrepLP} for how the symmetric eigenbasis representation of the bath hybridization can be obtained from a general non-symmetric representation.

\paragraph{Resolvent form.}
We can write $G(z)$ in Anderson form as the upper-left $N\times N$ block of the resolvent of
a matrix  $\HA \in \doubleC^{N(M'+1)\times N(M'+1)}$,
\begin{align}
    G(z) = \AND{A}_0 + \AND{B}_0 \left(\frac{1}{z - \HA}\right)_{1:N,\,1:N} \AND{B}_0 \, ,
\end{align}
where $\HA$ is a sparse matrix with the following block structure,
\begin{align}
    \label{eq:HAndDef}
    \HA =
    \begin{pmatrix}
        \AND{A}_1    & \AND{B}_1     & \cdots & \AND{B}_{M'}       \\
        \AND{B}_1    & \AND{a}_2 I_N & \cdots & 0                  \\
        \vdots       & \vdots        & \ddots & \vdots             \\
        \AND{B}_{M'} & 0             & \cdots & \AND{a}_{M'+1} I_N
    \end{pmatrix}.
\end{align}

\paragraph{Tight-binding interpretation.}

The matrix $\HA$ in \Eq{eq:HAndDef} can be interpreted as the one-particle representation of a non-interacting tight-binding Hamiltonian with one impurity site and \(M'\) bath sites. A convenient second-quantized form is
\begin{align}
    \label{eq:HAnd}
    \HA
     & =  \CR{i=1} \AND{A}_1 \AN{i=1} + \sum_{i=2}^{M'+1} \AND{a}_i \CR{i}\AN{i} \\  & \nonumber+ \sum_{i=2}^{M'+1}
    \CR{i=1} \AND{B}_{i-1} \AN{i} + \CR{i} \AND{B}_{i-1} \AN{i=1} \, ,
\end{align}
using the compact vector notation from \Eq{eq:crvecoverm} to suppress the summation over the local spin-orbital indices $\mi$.
The impurity Green's function for this Hamiltonian is
\begin{align}
    G(z)_{\mi,\mi'} =
    \brakettt{0}{\AN{\mi,i=1}\,\frac{1}{z-\HA}\,\CR{\mi',i=1}}{0} \, .
\end{align}

\paragraph{\label{sec:TriToAnd}Construction from the tridiagonal representation.}

We can construct an Anderson block representation from a tridiagonal block representation. The blocks $\AND{A}_0$ and $\AND{B}_0$ are the same in the Anderson representation as in the tridiagonal representation,
\begin{align}
    \AND{A}_0 = \TRI{A}_0 \, ,\qquad
    \AND{B}_0 = \TRI{B}_0 \,.
\end{align}
To obtain the remaining bath parameters, we start from the tridiagonal representation, i.e.\ from $\Ht$ [\Eq{eq:Htridiagonal}]. We construct a unitary matrix $U \in \doubleC^{NM \times NM}$ such that its upper-left $N\times N$ block is equal to $I_N$, while the remaining $N(M-1)\times N(M-1)$ block consists of the eigenvectors, written in column order, of the lower $N(M-1)\times N(M-1)$ block of $\Ht$. This transformation diagonalizes the bath subspace and yields
\begin{align}
    \label{eq:Heffanddiag}
    U^\dag \Ht U =
    \begin{pmatrix}
        \AND{A}_1               & \AND{b}_1 & \cdots & \AND{b}_{N(M-1)}   \\
        (\AND{b}_1)^\dag        & \AND{a}_2 & \cdots & 0                  \\
        \vdots                  & \vdots    & \ddots & \vdots             \\
        (\AND{b}_{N(M-1)})^\dag & 0         & \cdots & \AND{a}_{N(M-1)+1}
    \end{pmatrix},
\end{align}
where each $\AND{b}_i \in \doubleC^{N \times 1}$ is a column vector. In this form, the impurity couples to $N(M-1)$ bath levels, each with a single internal degree of freedom. To restore bath sites with internal dimension $N$, we define the Hermitian weight matrices as the outer product
\begin{align}
    \label{eq:WiA}
    \AND{W}_i = \AND{b}_i (\AND{b}_i)^{\dag} \, ,
\end{align}
and their Hermitian principal square roots
\begin{align}
    \label{eq:BiA}
    \AND{B}_i = (\AND{W}_i)^{1/2} \, .
\end{align}

With these blocks, we construct $\HA$ as in \Eq{eq:HAndDef}: the eigenvalues $\AND{a}_i$ of the bath subspace define the diagonal blocks $\AND{a}_i I_N$, and the coupling between the impurity and bath site $i$ is given by $\AND{B}_i$ from \Eqs{eq:WiA} and~\eqref{eq:BiA}.

Since $\AND{b}_i$ is a column vector, $\AND{W}_i=\AND{b}_i (\AND{b}_i)^\dagger$ is an outer product and therefore has $\mathrm{rank}(\AND{W}_i)\le 1$ (rank $0$ if the coupling vanishes). If several bath eigenvalues $\AND{a}_i$ are degenerate, one should recombine the corresponding couplings within the degenerate subspace. In that case, one may sum the weights $\AND{W}_j$ for all states with the same eigenvalue $\AND{a}_j$, $\AND{W}=\sum_j \AND{W}_j$, and obtain the impurity-bath coupling from the principal square root of this sum, $\AND{B}=\AND{W}^{1/2}$. This ensures a symmetry-conserving representation of the impurity-bath hybridization function.

\paragraph{Construction from the list-of-poles representation.}\label{sec:LP->And}

To obtain an Anderson representation of $G(z)$ from its list-of-poles representation,
we first-of-all find for the constant term
\begin{equation}
    \AND{A}_0 = A_0 \, .
\end{equation}
In order to determine the other block elements we start with the pole energies $\GPole{i}$ and block matrices $\GHyb{i}$ that define the list-of-pole representation \Eq{eq:GLPdef} and define the block-vector
\begin{equation}
    \BlockVector=
    \begin{pmatrix}
        \GHyb{1}    \\
        \GHyb{2}    \\
        \vdots \\
        \GHyb{M}
    \end{pmatrix}
    \in \doubleC^{NM \times N} \, ,
\end{equation}
and diagonal matrix
\begin{equation}
    H_{\mathrm{LP}} = \diag\!\left(\GPole{1} I_N,\, \GPole{2} I_N,\, \ldots,\, \GPole{M} I_N\right) \in \doubleR^{NM \times NM} \, .
\end{equation}
This allows us to rewrite our Green's function as
\begin{equation}
    \label{eq:GzResHe}
    G(z) = \AND{A}_0
    +
    \BlockVector^\dag \frac{1}{z - H_\mathrm{LP}} \BlockVector \, .
\end{equation}
As the column vectors in $\BlockVector$ are not orthonormal
for general response functions,
(they are for Green's functions),
we use the same symmetric orthonormalization as in \Appendix{sec:LP->Tri}
by using the overlap matrix
\begin{equation}
    S
    =
    \BlockVector^\dag \BlockVector
    =
    \sum_{i=1}^M \GHyb{i} \GHyb{i}
    =
    \sum_{i=1}^M \GWeight{i} \, ,
\end{equation}
to create the matrices
\begin{align}
    R_a       & = \BlockVector S^{-1/2} \in \doubleC^{NM \times N} \, , \nonumber \\
    \AND{B}_0 & = S^{1/2} \, ,
\end{align}
with $R_a^\dag R_a = I_N$.
We then find the orthogonal complement $R_\bara \in \doubleC^{NM \times N(M-1)}$
and collect the column vectors to generate the unitary matrix $U = R_a \oplus R_\bara \in \doubleC^{NM \times NM}$.

We then insert the identity, $U U^\dag$ twice, into \Eq{eq:GzResHe}, which yields
\begin{align}
    G(z)
     & =
    \AND{A}_0 + \AND{B}_0 R_a^\dag \frac{1}{z - U U^\dag H_\mathrm{LP} U U^\dag} R_a \AND{B}_0 \nonumber \\
     & =
    \AND{A}_0 + \AND{B}_0 R_a^\dag U \frac{1}{z - U^\dag H_\mathrm{LP} U} U^\dag R_a \AND{B}_0 \, .
\end{align}
We now use the property
\begin{align}
    U^\dag R_a
    =
    \begin{pmatrix}
        R_a^\dag \\
        R_\bara^\dag
    \end{pmatrix}
    R_a
    =
    \begin{pmatrix}
        I_{N \times N} \\
        0_{N(M-1) \times N}
    \end{pmatrix},
\end{align}
which can be interpreted as taking the upper left $N \times N$ block of
the resolvent of $U^\dag H_\mathrm{LP} U$
\begin{equation}
    G(z)
    =
    \AND{A}_0 + \AND{B}_0 \left[\frac{1}{z - U^\dag H_\mathrm{LP} U}\right]_{1:N,\,1:N} \AND{B}_0 \, .
\end{equation}

In the next step,
we want to bring the matrix $U^\dag H_\mathrm{LP} U$ into the form of \Eq{eq:Heffanddiag}.
The derivation for the Green's function in Anderson representation is then analogous
to the second step in transforming $G(z)$ from tridiagonal form to Anderson form.
After diagonalizing the lower right $N(M-1) \times N(M-1)$ block,
the Anderson matrix associated with $G(z)$ takes the form of \Eq{eq:Heffanddiag}.
To obtain the symmetric coupling matrices $\AND{B}_i$,
we follow the procedure of \Appendix{sec:TriToAnd} starting from \Eq{eq:Heffanddiag}.

\section{\label{sec:DiscreteG}Discrete self-energy starting from different representations.}

In the main text, we have found for the self-energy:
\begin{align}
    \label{eq:GzdownfoldedSelfEnergy2}
     & \Sigma(z) - \Sigma^\mH =  \left( \bigl[ \tG^{-1}(z) \bigl]_{22} \right)^{-1}                                            \\
    \nonumber
     & = S^{1/2} \frac{1}{ z - S^{1/2} \left[ \tS^{-1/2} \tH_{\mathrm{eff}}(z)  \tS^{-1/2} \right]_{22} S^{1/2} } S^{1/2} \, ,
\end{align}
where $\tG(z)$ is a $\tN \times \tN$ matrix-valued propagator with $\tN = 2N$. The matrix $\tS$ is the norm of $\tG(z)$ and
$S = \bigl( [ \tS^{-1}]_{22} \bigr)^{-1}$
is the norm of $\Sigma(z) - \Sigma^\mH$. \Eq{eq:GzdownfoldedSelfEnergy2} can be evaluated by
applying three steps to $\tG(z)$:
(i) inversion, (ii) projection to the $22$ block, and (iii) reinversion. In the following three subsections,
we show, using equations from the literature~\cite{lowdin_non-orthogonality_1950, lowdin_note_1951, lowdin_quantum_1955, andersen_linear_1975, andersen_muffin-tin_2000, gunnarsson_density-functional_1989, anisimov_density-functional_1991, koch_sum_2008,haverkort_multiplet_2012,lu_efficient_2014}, how these steps can be implemented explicitly, starting from any one of three
different representations of $\tG(z)$,
namely a tridiagonal, Anderson or list-of-pole representation.

\subsection{\label{sec:downfoldingTri}Starting from tridiagonal representation of \texorpdfstring{$\tG(z)$}{the propagator}}

We start from
\begin{align}
    \tG(z) =
    \TRI{\tB}_0^{1/2}
    \frac{1}{z - \TRI{\tA}_1 - \TRI{\tB}_1 \frac{1}{z - \TRI{\tA}_2 -  \hdots} \TRI{\tB}_1}
    \TRI{\tB}_0^{1/2} \, .
\end{align}
Comparing this expression to \Eq{eq:GzResHeff} gives
\begin{align}
    \tS^{1/2}             & = \TRI{\tB}^{1/2},                                                               \\
    \tH_{\mathrm{eff}}(z) & =  \TRI{\tA}_1 + \TRI{\tB}_1 \frac{1}{z - \TRI{\tA}_2 - \hdots} \TRI{\tB}_1 \, ,
\end{align}
with $\tH_{\mathrm{eff}}(z)$ also given in tridiagonal representation.

(i) \emph{Inversion.}
The inverse of $\tG(z)$ is written as
\begin{equation}
    \tG^{-1}(z)
    =
    \tS^{-1/2}
    \left(z - \TRI{\tA}_1 - \TRI{\tB}_1 \frac{1}{z - \TRI{\tA}_2 - \hdots} \TRI{\tB}_1\right)
    \tS^{-1/2} \, .
\end{equation}

(ii) \emph{Projection.}
To project to the lower-right block, we need to calculate
\begin{align}
    \left[ \tS^{-1/2} \tH_{\mathrm{eff}}(z)  \tS^{-1/2} \right]_{22} =  \left[ \tS^{-1/2} \TRI{\tA}_1  \tS^{-1/2} \right]_{22} & \nonumber \\ + \left[ \tS^{-1/2} \TRI{\tB}_1
        \frac{1}{z - \TRI{\tA}_2 - \hdots}
        \TRI{\tB}_1 \tS^{-1/2} \right]_{22} \, .
\end{align}

To bring the full result into a tridiagonal form with block dimension $N$, rather than retaining intermediate blocks of dimension $\tN$, one may apply a block-Lanczos procedure to
\begin{align}
    \tS^{-1/2}\TRI{\tB}_1
    \frac{1}{z - \TRI{\tA}_2 - \hdots}
    \TRI{\tB}_1\tS^{-1/2} \, ,
\end{align}
using as starting block-vector the projection onto the subspace $2$ at block $i=1$. With
\begin{align}
    S               & = \bigl( [ \tS^{-1}]_{22} \bigr)^{-1} \, , \nonumber                    \\
    \bar{\TRI{A}}_1 & = \left[ \tS^{-1/2} \TRI{\tA}_1  \tS^{-1/2} \right]_{22} \, , \nonumber \\
    \bar{\TRI{B}}_1 \frac{1}{z - \TRI{A}_2 - \hdots} \bar{\TRI{B}}_1
                    & = \nonumber                                                             \\
    \quad           & \left[ \tS^{-1/2} \TRI{\tB}_1
        \frac{1}{z - \TRI{\tA}_2 - \hdots}
        \TRI{\tB}_1 \tS^{-1/2} \right]_{22} \, ,
\end{align}
the projection of $\tG^{-1}(z)$ is written as
\begin{equation}
    \label{eq:GTrifolded}
    \bigl[\tG^{-1}(z) \bigr]_{22}
    =
    S^{-1} z
    - \bar{\TRI{A}}_1
    - \bar{\TRI{B}}_1 \frac{1}{z - \TRI{A}_2 - \hdots} \bar{\TRI{B}}_1 \, .
\end{equation}

(iii) \emph{Reinversion.}
To invert the previous expression, we first define
\begin{align}
    \TRI{A}_1 & = S^{1/2} \bar{\TRI{A}}_1 S^{1/2} \, , \nonumber \\
    \TRI{B}_1 & = S^{1/2} \bar{\TRI{B}}_1 \, ,
\end{align}
to rewrite \Eq{eq:GTrifolded} as
\begin{equation}
    \bigl[\tG^{-1}(z) \bigr]_{22}
    =
    S^{-1/2}
    \left(
    z
    - \TRI{A}_1
    - \TRI{B}_1 \frac{1}{z - \TRI{A}_2 - \hdots} \TRI{B}_1
    \right)
    S^{-1/2} \, .
\end{equation}
Inverting this gives us the self-energy in tridiagonal form
\begin{equation}
    \Sigma(z)
    =
    \Sigma^\mH
    +
    S^{1/2}
    \frac{1}{z - \TRI{A}_1 - \TRI{B}_1 \frac{1}{z - \TRI{A}_2 - \hdots} \TRI{B}_1}
    S^{1/2} \, .
\end{equation}

\subsection{\label{sec:downfoldingAnd}Starting from Anderson representation
    of \texorpdfstring{$\tG(z)$}{the propagator}}

For a response function \(\tG(z)\) in Anderson representation
\begin{align}
    \tG(z) = \tS^{1/2} \frac{1}{z - \AND{\tA}_1 - \sum_{i} \AND{\tB}_i \frac{1}{z-\AND{\ta}_{i+1}} \AND{\tB}_i} \tS^{1/2} \, ,
\end{align}

$\tH_{\mathrm{eff}}(z)$ is given in a list-of-poles representation:
\begin{align}
    \tH_{\mathrm{eff}}(z)  = \AND{\tA}_1 + \sum_{i} \AND{\tB}_i \frac{1}{z-\AND{\ta}_{i+1}} \AND{\tB}_i \, .
\end{align}

(i) \emph{Inversion.}
The inverse of  $\tG(z)$ is given  as
\begin{equation}
    \tG^{-1}(z)
    =
    \tS^{-1/2}
    \left(z - \AND{\tA}_1 - \sum_{i} \AND{\tB}_i \frac{1}{z-\AND{\ta}_{i+1}} \AND{\tB}_i\right)
    \tS^{-1/2} \, .
\end{equation}

(ii) \emph{Projection.}
To project to the lower right block,
we define
\begin{align}
    S               & = \bigl( [ \tS^{-1}]_{22} \bigr)^{-1} \, , \nonumber                                                 \\
    \bar{\AND{A}}_1 & = \left[\tS^{-1/2} \AND{\tA}_1\tS^{-1/2} \right]_{22} \, , \nonumber                                 \\
    \bar{\AND{B}}_i & = \left(\left[\tS^{-1/2} \AND{\tB}_i \AND{\tB}_i \tS^{-1/2} \right]_{22}\right)^{1/2} \, , \nonumber \\
    \AND{a}_i       & = \AND{\ta}_i \, ,
\end{align}
and express the projection of $\tG^{-1}(z)$ as
\begin{equation}
    \label{eq:GAndfolded}
    \bigl[\tG^{-1}(z) \bigr]_{22}
    =
    S^{-1}z - \bar{\AND{A}} - \sum_{i} \bar{\AND{B}}_i \frac{1}{z-\AND{a}_{i+1}} \bar{\AND{B}}_i \, .
\end{equation}

(iii) \emph{Reinversion.}
To invert the previous expression,
we first define
\begin{align}
    \AND{A}_1 & = S^{1/2} \bar{\AND{A}}_1 S^{1/2}\, , \nonumber                           \\
    \AND{B}_i & = \left(S^{1/2} \bar{\AND{B}}_i \bar{\AND{B}}_i S^{1/2}\right)^{1/2} \, ,
\end{align}
to rewrite \Eq{eq:GAndfolded} as
\begin{equation}
    \bigl[\tG^{-1}(z) \bigr]_{22}
    =
    S^{-1/2}
    \left(
    z
    - \AND{A}_1
    - \sum_i \AND{B}_i \frac{1}{z - \AND{a}_{i+1}} \AND{B}_i
    \right)
    S^{-1/2} \, .
\end{equation}
Inverting this gives us the self-energy in Anderson form
\begin{align}
    \Sigma(z) = \Sigma^\mH + S^{1/2} \frac{1}{z - \AND{A}_1 - \sum_{i} \AND{B}_i \frac{1}{z-\AND{a}_{i+1}} \AND{B}_i} S^{1/2} \, .
\end{align}

\subsection{\label{sec:downfoldingLP}Starting from list-of-poles representation of \texorpdfstring{$\tG(z)$}{the propagator}}

In this appendix, we describe how to obtain the pole expansion of the self-energy $\Sigma(z)$ from the pole expansion of $\tG(z)$.
If the transformations between different types of response functions (\Appendices{sec:LP->Tri},~\ref{sec:LP->And}) are available, one could first transform $\tG(z)$ to tridiagonal or Anderson representation and continue as described in \Appendices{sec:downfoldingTri} or~\ref{sec:downfoldingAnd}, respectively, and transform the resulting $\Sigma(z)$ back to a list-of-poles representation at the end.

Below, we describe a self-contained variant of this route: starting from the list-of-poles representation of $\tG(z)$, we bring it into Anderson form, perform the three steps (i) inversion, (ii) projection, and (iii) reinversion, and obtain the self-energy $\Sigma(z)$ again in a list-of-poles representation.
This is the approach used to obtain the TaSK and NRG results shown in \Figs{fig:fit_SEdisc} and~\ref{fig:SIAM_lowFrequency}.

We consider a matrix-valued function $\tG(z)
    \in \doubleC^{\tN \times \tN}$ with elements $[\tG(z)]_{\MatElOne \MatElTwo}$ and
$\MatElOne = 1, \ldots,\tN=2N$,  as defined
in \Sec{sec:SigmaFromDiscreteEQM} of the main text: $\MatElOne = 1,\dots, N$
enumerates $a^\pdag_{0,\mi}$, $a^\dagger_{0,\mi}$ operators,
$\MatElOne = N+1,\dots, 2N$ enumerates
the corresponding $q^\pdag_{\mi}$,
$q_\mi^\dagger$ operators.
We assume $\tG(z)$
to be given as a list-of-pole representation,
\begin{align}
    \label{eq:tGz_discrete}
    \tG(z) = \sum_{i=1}^M \frac{\AugGWeight{i}}{z - \AugGPole{i}} \, ,
\end{align}
(see \Appendix{sec:notation} for notation)
having three key properties, all of which are used below: (i) The $M$ poles are real, $\AugGPole{i} \in \doubleR$; (ii) the $M$ matrices $\AugGWeight{i} \in \doubleC^{\tN \times \tN}$ are  Hermitian and positive semidefinite;
and (iii) each $\AugGWeight{i}$ matrix
has rank one (this does not imply loss of generality, see \Appendix{sec:symeigrepLP}).  Then, each $\AugGWeight{i}$ can be written as the outer product, $\AugGWeight{i} = \AugGHyb{i}^{\dag} \AugGHyb{i}$, of a row vector $\AugGHyb{i}^\pdag \in
    \doubleC^{1 \times \tN}$ and its Hermitian conjugate column vector $\AugGHyb{i}^\dagger \in \doubleC^{\tN \times 1}$, with elements  $[\AugGWeight{i}]_{\MatElOne \MatElTwo} = [\AugGHyb{i}^{\dag}]_{\MatElOne} [\AugGHyb{i}]_{\MatElTwo}$.

We will call the collection of data $\{\AugGPole{i}, \AugGWeight{i}\}$ the spectrum of $\tG(z)$; it describes the positions and matrix-valued weights of the discrete $\delta$-peaks of $-\frac{1}{\pi}
    \mathrm{Im} \tG(\omega+\mr{i}0^+)$.  The spectrum is fully encoded in the two matrices
$\{\tOmega, \AugBlockVector\}$, where $\tOmega  = \diag(\AugGPole{1}, \dots, \AugGPole{M}) \! \in \! \doubleR^{M\times M}$ is diagonal
and $\AugBlockVector \in  \doubleC^{M\times \tN}$
has elements
$\AugBlockVector_{i\MatElOne} = [\AugGHyb{i}]_{\MatElOne}$. Indeed, \begin{align}
    \tG(z) =  \AugBlockVector^{\dag} \frac{1}{z - \tOmega} \AugBlockVector \, .
\end{align}
The integrated weight matrix $\tS$
of
\Eq{eq:DefStilde} is given by
\begin{align}
    \tS = \AugBlockVector^{\dag} \AugBlockVector \in \doubleC^{\tN \times \tN} \, .
\end{align}
We can further perform the polar decomposition~\footnote{
    This decomposition
    can be found via a singular value decomposition, \unexpanded{$\AugBlockVector= U S V^\dagger = (U V^{\dagger}) (V S V^{\dagger}) = \tR_a \tS^{1/2}$},
    which avoids inversion of \unexpanded{$\tS^{1/2}$}.}
\begin{align}
    \label{eq:PolarDecomposition}
    \AugBlockVector = \tR_a \tS^{1/2} \, , \quad \tR^{\dag}_a \tR_a = I_{\tN} \, ,
\end{align}
where $\tR_a \in \doubleC^{M \times \tN}$ is an isometry and $\tS^{1/2} \in \doubleC^{\tN \times \tN}$
is Hermitian and positive semidefinite. Then, the latter is the principal square root of $\tS$, justifying the notation $\tS^{1/2}$. (Note: the subscript on $\tR_a$ is a single-valued label, not an index, used to distinguish $\tR_a$ from $\tR_\bara$, defined below.) Furthermore,
the propagator $\tG(z)$ takes the form
\begin{align}
    \label{eq:tG_TR}
    \tG(z) = \tS^{1/2} \tR^{\dag}_a \frac{1}{z - \tOmega} \tR_a \tS^{1/2} \, ,
\end{align}
with the norm factors $\tS^{1/2}$ split from the dynamical part.

(i) \textit{Inversion}. To obtain a DSR for $G^{-1}(z)$,  we seek to transcribe $G(z)$ to the form
\begin{equation}
    \label{eq:tG_inv}
    \tG(z) = \tS^{1/2} \frac{1}{z - \tH_\mathrm{eff}(z)} \tS^{1/2}
\end{equation}
and find the DSR of $\tH_\mathrm{eff}(z) \in \doubleC^{\tN \times \tN}$.
To this end, our strategy will be to express $\tG(z)$ as the projection of the $M \times M$ matrix $(z - \tOmega)^{-1}$ to an $\tN \times \tN$ block submatrix, and then to invert the latter using the Schur complement formula, which involves other blocks of $\tOmega$.

For that, we define an orthogonal complement to $\tR_a$, the isometry
$\tR_{\bara} \in \doubleC^{M \times \bN}$, with $\bN = M - \tN$,
such that $\tR^\dag_a \tR^{\pdag}_{\bara} = 0$ and
$\tR^\dag_\bara \tR^{\pdag}_{\bara} = I_\bN$%
~\footnote{
    To find $\protect\tR_\bara$,  we construct
    $P_a =  \protect\tR^\pdag_a \protect\tR_a^{\dagger}$
    the projector leaving
    $\protect\tR_a$ invariant, QR-decompose its complement,   $Q R = P_\bara = I_M - P_a  $, and set
    $\protect\tR_\bara = Q$.}.
Then, $\tU = \mbox{$\tR_a \oplus \tR_\bara$} \in \doubleC^{M \times M}$, built
from the columns of $\tR_a$ and $\tR_\bara$, is unitary. Next, we define the blocks $\tOmega_{x\tilde{x}} = \tR^\dag_{x} \tOmega \tR^{\pdag}_{\tilde{x}}$, for $x,\tilde{x} \in \{a,\bara\}$, and write
\begin{flalign}
    \label{eq:block-form-for-Omega}
                                                & \begin{array}{c}\tOmega    = (\tR_a , \tR_\bara) \\ \phantom{\tR} \end{array}
    \! \! \begin{pmatrix}
              \tOmega_{aa}      & \!\tOmega_{a \bara}     \\
              \tOmega_{\bara a} & \!\tOmega_{\bara \bara}
          \end{pmatrix}
    \! \begin{pmatrix}
           \tR^{\dag}_a \\ \tR^{\dag}_\bara
       \end{pmatrix}\!
    \begin{array}{c} = \tU \\ \phantom{\tR} \end{array}
    \!\!\! \begin{pmatrix}
               \tOmega_{aa}      & \!\tOmega_{a \bara}     \\
               \tOmega_{\bara a} & \!\tOmega_{\bara \bara}
           \end{pmatrix} \!\!
    \begin{array}{c} \tU^{\dag}
        \\ \phantom{R}\end{array} ,   \hspace{-5mm} &
\end{flalign}
obtaining a block decomposition of $\tOmega$
in the basis of the columns of $\tU^{\dagger}$.
With this notation, \Eq{eq:tG_TR} yields
\begin{align}
    \label{eq:tG=aaBlockofInverse}
    \tG(z) & =
    \tS^{1/2} \Bigl(\frac{1}{z-\tOmega}\Bigr)_{aa} \tS^{1/2} \, .
\end{align}
$\tG(z)$ thus involves the projection of the inverse of a $2\times 2$ block matrix to
its $aa$ block.
This calls for use of the Schur complement formula for block matrix inversion,
{%
        \addtolength{\belowdisplayskip}{-12pt}%
        \addtolength{\belowdisplayshortskip}{-12pt}%
        \begin{align}
            \label{eq:Schur-Supp}
            \left( \!
            \begin{pmatrix}
                M_{aa} & M_{a \bara} \\ M_{\bara a} & M_{\bara \bara}
            \end{pmatrix}^{\!\! -1} \right)_{\!\! aa} =  \Bigl( M_{aa} - M_{a \bara}
            \frac{1}{M_{\bara \bara}} M_{\bara a} \Bigr)^{-1} ,
        \end{align}}%
which yields
\begin{align}
    \label{eq:tG^-1Adag[]A}
    \Bigl(\frac{1}{z-\tOmega}\Bigr)_{aa} & =
    \Bigl(z-\tOmega_{aa} - \tOmega_{a \bara}
    \frac{1}{z-\tOmega_{\bara \bara}} \tOmega_{\bara a}
    \Bigr)^{-1}  \, .
\end{align}
By comparing \Eq{eq:tG_inv} to
\Eqs{eq:tG=aaBlockofInverse} and~\eqref{eq:tG^-1Adag[]A}, we find
\begin{align}
    \label{eq:explicitSigma-a}
    \tH_{\mathrm{eff}}(z) &
    =\tOmega_{aa}+ \tOmega_{a \bara}
    \frac{1}{z-\tOmega_{\bara \bara}} \tOmega_{\bara a}  \, .
\end{align}
Now we diagonalize the block $\tOmega_{\bara \bara}$
as $\bU^\dag \, \tOmega_{\bara \bara} \bU^\pdag =  {\bOmega}$,
where ${\bOmega} = \diag(\HeffPole{1}, \dots, \HeffPole{\bN}) \in \doubleR^{\bN \times \bN}$ is diagonal and $\bU \in \doubleC^{\bN \times \bN}$ is unitary, and define $\BlockVectorHeff = \bU^{\dag} \tOmega_{\bara a}  \in \doubleC^{\bN \times \tN}$. Then,  \Eq{eq:explicitSigma-a} leads to the desired DSR for $\tH_{\mathrm{eff}}(z)$,
\begin{align}
    \label{eq:S_spectral}
    \tH_{\mathrm{eff}}(z) = \tOmega_{aa} +  \sum_{i=1}^{\bN} \frac{\HeffWeight{i}}{z - \HeffPole{i}}  \, ,
\end{align}
(see \Appendix{sec:notation} for notation)
where each $\HeffWeight{i} \in \doubleC^{\tN \times \tN}$
is a rank-1 matrix with elements given by
$[\HeffWeight{i}]_{\MatElOne \MatElTwo} =
    [\BlockVectorHeff^{\dag}]_{\MatElOne i }
        [\BlockVectorHeff]_{i \MatElTwo} $.
At this point, we have transformed $\tG(z)$ from the list-of-poles to the Anderson representation, which allows us to invert it straightforwardly:
\begin{align}
    \tG^{-1} (z) = \tS^{-1/2} \Biggl[z -
        \tOmega_{aa} \mathbin{-}
        \sum_{i=1}^{\bN} \frac{\HeffWeight{i}}{z - \HeffPole{i}}
        \Biggr] \tS^{-1/2} \, .
\end{align}

(ii) \emph{Projection}.
The self-energy formula \Eq{eq:GzdownfoldedSelfEnergy2}
involves $\bigl[\tG^{-1}(z)\bigr]_{22} \in \doubleC^{N\times N}$,
with $N=\tN/2$. This is the projection of $\tG^{-1}(z)$, viewed as
a $2 \!\times \! 2$ block matrix of four blocks
$\in \doubleC^{N \times N}$, to its 22 block.
We express it as
\begin{align}
    \bigl[\tG^{-1} (z)\bigr]_{22} & = S^{-1/2} \Biggl[z -
        \Omega_{aa} \mathbin{-}
        \sum_{i=1}^{\bN} \frac{W_i}{z - \HeffPole{i}}
        \Biggr] S^{-1/2} \, ,
\end{align}
where  we defined
$S = \bigl( [ \tS^{-1} ]_{22} \bigr)^{-1} \in \doubleC^{N\times N}$,
denoted its principal square root by $S^{1/2}$, and defined the matrices
\begin{subequations}
    \begin{align}
        \Omega_{aa} & = S^{1/2} \bigl[\tS^{-1/2}\tOmega_{aa}\tS^{-1/2} \bigr]_{22} S^{1/2}
        \, \in \doubleC^{N\times N}, \nonumber
        \\
        W_i         & =  S^{1/2} \bigl[ \tS^{-1/2} \HeffWeight{i} \tS^{-1/2} \bigr]_{22} S^{1/2}
        \, \in \doubleC^{N\times N} .
    \end{align}
\end{subequations}
Since all $\HeffWeight{i}$ are Hermitian rank-1 matrices, the same is true for all $W_i$, hence their matrix elements can be expressed as
$[W_i]_{\MatElOne \MatElTwo} = [\BlockVector^\dagger]_{\MatElOne i} [\BlockVector]_{i \MatElTwo}$,
with $\BlockVector \in \doubleC^{\bN \times N}$.
By \Eq{eq:GzdownfoldedSelfEnergy2}, the self-energy
can thus be expressed as
\begin{equation}
    \label{eq:Sigma-projected}
    \Sigma(z) -
    \Sigma^\mH
    =  S^{1/2}
    \Bigl(z - \Omega_{aa} - \BlockVector^\dagger \frac{1}{z -
        \peakbar{\Omega}} \BlockVector \Bigr)^{-1} \!\!
    S^{1/2} \, .
\end{equation}

(iii) \emph{Reinversion}.
Our final goal is to obtain a list-of-poles representation for the self-energy from \Eq{eq:Sigma-projected}. To this end,
we use the Schur complement formula~\eqref{eq:Schur-Supp}
to express the central factor of~\eqref{eq:Sigma-projected} as
[see \Eq{eq:tG^-1Adag[]A}]
\begin{align}
    \label{eq:Schur-backwards}
    \Bigl(z - \Omega_{aa} - \BlockVector^\dagger \frac{1}{z -
        \peakbar{\Omega}} \BlockVector \Bigr)^{-1} =
    \Bigl( \frac{1}{z - \Omega'} \Bigr)_{aa} ,
\end{align}
where we defined  the matrix
\begin{align}
    \label{eq:defineOmega'}
    \Omega' =
    \begin{pmatrix}
        \Omega_{aa} \; & \BlockVector^{\dag} \\
        \BlockVector              & \bOmega
    \end{pmatrix} \in \doubleC^{\bM \times \bM} ,
\end{align}
with $\bM \!=\!  N + \bN \!=\! M - N $.
We diagonalize it as $U'^\dagger \Omega' U' = \OmegaSigma$, where $\OmegaSigma = \diag(\SigmaPole{1},\dots,\SigmaPole{\bM}) \in \doubleR^{\bM \times \bM}$. Then
\begin{align}
    \label{eq:Schur-backwards-aa}
    \Bigl( \frac{1}{z - \Omega'} \Bigr)_{aa}
    = \Bigl( U' \frac{1}{z - \OmegaSigma} U^{\prime \dagger} \Bigr)_{aa}
    = R^{\prime \dagger}_a \frac{1}{z - \OmegaSigma} R'_a \, ,
\end{align}
where we collected the first $N$ columns of $U'^{\dag}$ to construct the isometry $R'_a \in \doubleC^{\bM \times N}$, with $R^{\prime \dagger}_a R'_a = I_N$.
Equations~\eqref{eq:Sigma-projected},~\eqref{eq:Schur-backwards}
and~\eqref{eq:Schur-backwards-aa}
yield the desired DSR for $\Sigma(z)$,
\begin{subequations}
    \begin{align}
        \Sigma(z) - \Sigma^{\mH}
         & =  {\BlockVectorSigma}^{\dag} \frac{1}{z-\OmegaSigma} \BlockVectorSigma
        = \sum_{i=1}^{\bM} \frac{\SigmaWeight{i}}{z - \SigmaPole{i}} \, ,
    \end{align}
\end{subequations}
where  $\BlockVectorSigma = R'_a S^{1/2} \in \doubleC^{\bM \times N}$, and each
$\SigmaWeight{i} \in \doubleC^{N \times N}$ is a  rank-1 matrix with elements $[\SigmaWeight{i}]_{\MatElOne \MatElTwo} = [{\BlockVectorSigma}^\dagger]_{i\MatElOne} [\BlockVectorSigma]_{i\MatElTwo}$.
Note that the norm of $\Sigma(z) - \Sigma^{\mH}$ is given by ${\BlockVectorSigma}^{ \dagger} \BlockVectorSigma = S$.

\section{\label{sec:update-impurity-problem}
  Update of the hybridization function using a discrete self-energy}

In the DMFT self-consistency loop, the hybridization function of the impurity model is updated using the impurity self-energy. In \RAS{} and \Quanty{} this update is performed directly at the level of the discrete pole representation of $\Delta(z)$ and $\Sigma(z)$. The effective hybridization function is obtained from~\cite{lu_efficient_2014}
\begin{align}
    \Delta^{\text{eff}}(z) = \Delta(z I_N-\Sigma(z)) \, ,
\end{align}
whereby the matrix valued arguments of $\Delta$ should be interpreted using the discrete representation of the hybridization function [\Eq{eq:DeltaHybDef}] as
\begin{align}
    \Delta^{\text{eff}}(z) = \sum_{i=1}^{M}V^\dagger_i\frac{1}{z-E_i - \Sigma(z)}V_i \, .
\end{align}
Note that this equation requires the spin-orbital basis of $\Delta(z)$ and $\Sigma(z)$ to be the same and independent of $z$, which for the examples in the current paper is trivially fulfilled.

For a self-energy in the list-of-poles representation, one obtains $\Delta^{\text{eff}}(z)$ as a sum over $M$ Anderson-type response functions:

\begin{align}\label{eq:UpdateDeltaSumOverAnderson}
    \Delta^{\text{eff}}(z) = \sum_{i=1}^{M}V^\dagger_i\frac{1}{z-E_i - \Sigma^{\mH} - \sum_{j=1}^{M^\Sigma}  \SigmaWeight{j} \frac{1}{z - \SigmaPole{j}}}V_i \, .
\end{align}

To bring $\Delta^{\text{eff}}(z)$ to the list-of-poles representation, we transform each summand in \Eq{eq:UpdateDeltaSumOverAnderson} from Anderson-type into the list-of-poles representation.

\begin{align}
    \Delta^{\text{eff}}(z) =  \sum_{i=1}^{M}  \left[\sum_{j=1}^{M^\Sigma +1} {V_j'}^\dagger \frac{1}{z - E'_j}V'_j \right]_i.
\end{align}
The double sum can be combined into a single sum over $M(M^\Sigma+1)$ poles:
\begin{align}
    \Delta^{\text{eff}}(z) = \sum_{i=1}^{M(M^{\Sigma}+1)} {V^{\text{eff}}_i}^{\dagger} \frac{1}{z - E^{\text{eff}}_i}V^{\text{eff}}_i \, .
\end{align}
The number of poles can then be reduced to keep the computational cost moderate.

\section{\label{sec:quasparticle-weight}Regularization of the quasiparticle weight from a discrete spectral representation}

In the main text we have shown how the quasiparticle weight $Z$ can be obtained from the DSR of the self-energy $\Sigma(z)$:
\begin{subequations}\label{eq:Z_from_Sigma_DSR}
    \begin{align}
        \Sigma(\omega + \I0^+) & = \Sigma^\mH + \sum_{i} \frac{\SigmaWeight{i}}{\omega + \I0^+ - \SigmaPole{i}} \, ,
        \\
        \partial_{\omega} \mr{Re} \, \Sigma(\omega + \I0^+) |_{\omega = 0}
                               & = - \sum_{\SigmaPole{i} \neq 0} \frac{\SigmaWeight{i}}{(\SigmaPole{i})^2} + \sum_{\SigmaPole{i} = 0} \frac{\SigmaWeight{i}}{(0^+)^2} \, ,
        \\
        Z                      & = \bigl[I_N - \partial_{\omega} \mr{Re} \, \Sigma(\omega + \I0^+) |_{\omega = 0}\bigr]^{-1} \, .
    \end{align}
\end{subequations}

Numerical algorithms might yield spurious poles with small ($\Tr(\SigmaWeight{i})\ll 1$) residue.
When such a pole occurs at small energies ($|\SigmaPole{i}| \sim \Tr(\SigmaWeight{i})$),
these spurious poles will dominate the calculation of the quasiparticle weight and may lead to $Z=0$ even for a metallic solution.
In this appendix we discuss strategies to regularize the calculation of $Z$ in the presence of such spurious poles.

A first approach is to remove poles whose spectral weight falls below a chosen threshold.
If the spurious poles are close to machine precision $(\Tr(\SigmaWeight{i}) < 10^{-14})$, which is the case for the application in \Sec{sec:Application1} of the main text, spurious and physical poles are well separated. In this case, $Z$ can be straightforwardly regularized by removing these poles.

A second approach becomes necessary when the spurious poles have larger spectral weight $(\Tr(\SigmaWeight{i}) \approx 10^{-7})$, which happens for the application in \Sec{sec:Application2} of the main text.
By choosing larger threshold values, one might risk removing poles that are physically relevant.
A more systematic way to regularize the calculation of $Z$ is to modify the expression for $Z$ in \Eq{eq:Z_from_Sigma_DSR} by introducing a regularization parameter $\lambda$:
\begin{align}
    \label{eq:quasiparticle-weight_stable}
    Z_{\lambda} = \left( I_N + \sum_i \SigmaWeight{i}  \frac{1}{ {\SigmaPole{i}}^2 + \lambda^2}  \right)^{-1}.
\end{align}
In the absence of spurious poles at the Fermi level $Z_{\lambda}$ will approach $Z$ when $\lambda \to 0$.
When spurious poles are present, too small $\lambda$ will lead to numerical instabilities, while too large $\lambda$ will modify the results.
Instead, one needs to find a range of $\lambda$ where $Z_{\lambda}$ is approximately constant with respect to $\lambda$.

\begin{figure}[tb]
    \includegraphics[scale=1]{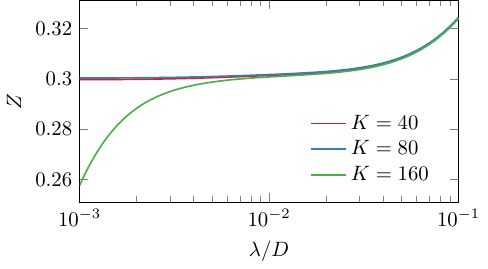}
    \caption{
        \label{fig:quasiparticle-weight-regularization}
        Quasiparticle weight $Z$ as a function of the regularization parameter $\lambda$
        at an interaction $U/D=2$.
        The parameter $K$ is the size of the Krylov space of the propagator $\tG$
        and indirectly determines the number of poles of the self-energy.
    }
\end{figure}

In \Fig{fig:quasiparticle-weight-regularization} we show the effect of
such a regularization on the self-energy represented by different number of poles. If no spurious poles are present ($K=40$ and $K=80$), one can take the limit $\lambda\to0$ without problems. For $K=160$ however, one needs to find a range where $Z_\lambda$ is approximately constant. In this case, it happens around $\lambda/D \approx 10^{-2}$, where $Z_{\lambda}$ has an inflection point. For intermediate values of $\lambda$, the quasiparticle weight is the same between the different Krylov sizes $K$, which determine the number of poles in $\Sigma(z)$.

\bibliography{references.bib}

\clearpage

\end{document}